\documentclass[aps,prb,reprint,floatfix]{revtex4-2}
\pdfoutput=1
\usepackage[utf8]{inputenc}
\usepackage[english]{babel}
\usepackage{microtype}
\usepackage{amsmath,amssymb,amsfonts}
\usepackage{braket}
\usepackage{bm}
\usepackage{graphicx}

\usepackage[
    pdftitle={Theory of driven Higgs oscillations
    and third-harmonic generation in unconventional superconductors},
    pdfauthor={Lukas Schwarz, Dirk Manske},
    colorlinks=true,
    unicode=true,
    pdfborder={0 0 0},
    allcolors=blue
]{hyperref}

\renewcommand{\vec}[1]{\bm{#1}}
\newcommand{\im}{\mathrm{i}}
\newcommand{\e}{\mathrm{e}}
\newcommand{\vk}{{\vec k}}
\newcommand{\VV}[2]{\begin{pmatrix}#1\\#2\end{pmatrix}}
\newcommand{\VVV}[3]{\begin{pmatrix}#1\\#2\\#3\end{pmatrix}}

\begin{document}

\title{Theory of driven Higgs oscillations and third-harmonic generation\\in
unconventional superconductors}

\author{Lukas Schwarz}

\author{Dirk Manske}

\affiliation{Max Planck Institute for Solid State Research,
70569 Stuttgart, Germany}

\date{\today}

\begin{abstract}
Higgs spectroscopy is a new field
in which Higgs modes in nonequilibrium superconductors are analyzed
to gain information about the ground state.
One experimental setup
in which the Higgs mode in $s$-wave superconductors was observed
is periodic driving with THz light,
which shows resonances in the third-harmonic generation (THG) signal
if twice the driving frequency matches the energy of the Higgs mode.
We derive expressions of the driven gap oscillations
for arbitrary gap symmetry and calculate the THG response.
We demonstrate that the possible Higgs modes
for superconductors with non-trivial gap symmetry
can lead to additional resonances
if twice the driving frequency matches the energy of these Higgs modes
and we disentangle the influence of charge density fluctuations (CDF)
to the THG signal within our clean-limit analysis.
With this we show
that THG experiments on unconventional superconductors
allow for a detection of their Higgs modes.
This paves the way for future studies on realistic systems
including additional features
to understand the collective excitation spectra
of unconventional superconductors.
\end{abstract}

\maketitle

\section{Introduction}
In recent years with the emergence of THz spectroscopy,
studies on matter were possible in regimes
which were inaccessible before
\cite{PhysicsToday.65.44,AdvOptPhoton.8.401,AdvPhys.65.58}.
Due to the energy of THz radiation in the meV range,
gentle excitations of materials can be performed
without destroying the quantum coherence of the whole system.
This allows for controlled experiments in nonequilibrium situations
from which ground state properties can be uncovered.

\begin{figure}[t]
    \includegraphics[width=0.5\textwidth]{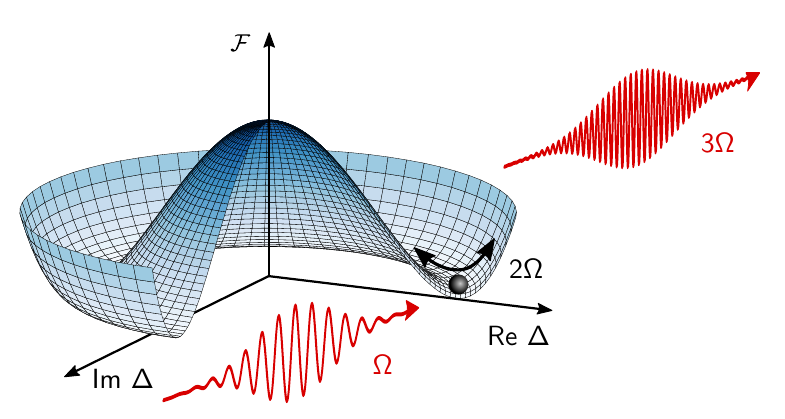}
    \caption{Free energy $\mathcal F$ of a superconductor
    as function of the complex order parameter $\Delta$
    shaped as a Mexican hat.
    Radial oscillations in the potential
    correspond to amplitude (Higgs) oscillations of the order parameter.
    If the superconductor is driven by light of frequency $\Omega$,
    enforced Higgs oscillations occur with a frequency of $2\Omega$
    due to the quadratic coupling of light.
    The system responds with third-harmonic generation (THG) $3\Omega$,
    resulting from the $\Omega$ drive plus $2\Omega$ Higgs oscillations.}
    \label{fig:mexican_hat}
\end{figure}

One interesting field within the THz studies
are excitations of collective modes like the Higgs mode in superconductors
\cite{PhysRevLett.13.508,JLowTempPhys.126.901,PhysRevB.84.174522,%
PhysRevB.87.054503,Science.345.1121,AnnuRevCondensMatterPhys.6.269}.
In equilibrium this mode arises
due to the spontaneous $U(1)$ symmetry breaking in the superconducting state
which is characterized by a Mexican hat-shaped free energy potential
(Fig.~\ref{fig:mexican_hat}).
In principle there are two different collective modes possible:
A phase excitation azimuthally around the brim of the Mexican hat
-- the Goldstone mode
and an amplitude excitation radially
-- the Higgs mode.
However, due to the Anderson-Higgs mechanism \cite{PhysRev.110.827},
the originally massless Goldstone mode gets shifted to the plasma energy,
whereas the Higgs mode remains with an energy of $2\Delta$,
i.e. the energy of the superconducting gap \cite{JLowTempPhys.126.901}.

The Higgs mode does not possess a dipole moment
therefore it does not couple linearly to light.
This makes its experimental excitation and detection difficult.
Except in the special situation,
where a charge-density wave order coexist
with the superconducting order on a similar energy range
providing a coupling between the two orders
and making the Higgs mode Raman-active
\cite{PhysRevLett.45.660,PhysRevLett.47.811,PhysRevB.26.4883,%
PhysRevB.89.060503,PhysRevB.97.094502,PhysRevLett.122.127001},
more effort has to be expended for an observation of the Higgs mode.
Only recently with ultrafast THz laser technology
an excitation and observation of the Higgs mode in a pump-probe experiment
for the $s$-wave system NbN and Nb$_{1-x}$Ti$_x$N was possible
\cite{PhysRevLett.109.187002,PhysRevLett.111.057002}
and first experiments on cuprate $d$-wave systems were reported
\cite{PNAS.110.4539,PhysRevLett.120.117001}.
In these experiments, oscillations of the optical conductivity
after a short THz pulse
resulting from Higgs oscillations in nonequilibrium could be observed.
Alternatively, it was shown theoretically
that time-resolved ARPES experiments could be be used
in a pump-probe setup as well for the observation of Higgs oscillations
\cite{PhysRevB.92.224517,PhysRevB.96.184518,arxiv.2002.05904}.
These oscillations in nonequilibrium
were predicted already before the experiment
\cite{SovPhysJETP.38.1018,PhysRevLett.93.160401,PhysRevLett.96.097005,%
PhysRevLett.96.230403,PhysRevLett.96.230404,PhysRevB.76.224522,%
PhysRevB.77.180509,PhysRevB.78.132505,PhysRevB.95.104507}
and lead to a great variety of further research
considering coupling to phonons \cite{PhysRevB.84.214513,PhysRevB.90.014515},
quasi-1d systems \cite{NewJPhys.15.055016},
multiband superconductors with additional Leggett modes
\cite{EurophysLett.101.17002,NatCommun.7.11921},
strongly-coupled regimes beyond BCS theory \cite{PhysRevB.93.094509},
current-carrying states \cite{PhysRevLett.118.047001,PhysRevLett.122.257001}
and condensates with $d$-wave symmetry
\cite{PhysRevLett.115.257001,NatCommun.11.287}.

Especially the consideration of unconventional superconductors
may open a new field of spectroscopy.
For superconductors with non-trivial gap symmetry,
there are multiple Higgs modes possible
which can be decomposed into the irreducible representation
of the underlying lattice symmetry \cite{PhysRevB.87.054503}
and can be understood as asymmetric oscillations of the condensate.
Therefore one may infer the symmetry of the gap
from a careful analysis of the Higgs modes in a chosen system
\cite{NatCommun.11.287}.

Recent studies on light-induced superconductivity in different systems
\cite{Science.331.189,PhysRevB.89.184516,Nature.7591.461,%
JPhysSocJap.88.044704,arxiv.1905.08638,arxiv.1908.10879}
raise the question on how to define superconductivity
in a short-lived nonequilibrium state.
One criteria for superconductivity is the Meissner effect,
which is induced on a microscopic level by the Anderson-Higgs mechanism.
Hence, a measurement of the Higgs mode is an equivalent probe
and it is therefore important to understand Higgs oscillations
for all kind of unconventional superconductors.

As pump-probe experiments are difficult to conduct
due to the requirement of ultrashort single-cycle THz pulses
within the energy range of the superconducting gap,
it is worth to consider alternative experimental setups
for measuring Higgs oscillations.
Instead of quenching the superconducting condensate with a THz pump pulse
and observing the intrinsic Higgs oscillations directly,
a periodic driving scheme with multicycle pulses can be used
(Fig.~\ref{fig:mexican_hat}).
The nonlinear coupling to the condensate
of light $A(t)$ with frequency $\Omega$
induces oscillations of the energy gap $\delta\Delta(t)$
with twice the driving frequency $2\Omega$.
This nonlinear coupling also induces higher-order currents $j$,
where a third-harmonic generation (THG) component $j^{(3)}(t)$ arises,
resulting from the driven gap oscillations,
i.e. $j^{(3)}(t) \propto A(t) \delta\Delta(t)$
\cite{PhysRevB.92.064508}.
If the effective $2\Omega$ driving frequency
is tuned to the energy of the Higgs mode $2\Delta$,
a resonance occur in the gap oscillation
and consequently also in the THG intensity,
which indicates the existence of the Higgs mode.
This effect can be measured in the transmitted electric field.

This setup was already successfully demonstrated
in an experiment for the $s$-wave superconductor NbN
\cite{Science.345.1145,PhysRevB.96.020505}.
More recently, experiments on several different cuprates were performed,
where a possible new mode at an energy below the symmetric $2\Delta$ Higgs mode
was observed in the THG spectrum \cite{arxiv.1901.06675}.
This experiment and further ongoing efforts
in performing these kinds of experiments on unconventional superconductors
require an understanding of the underlying physics.
Hence, this work investigates the THG response
for different unconventional superconductors
to study the influence of gap symmetries.

There are several experimental difficulties while performing THG experiments.
With current technology, the frequency of the driving field in the THz range
cannot be tuned arbitrarily.
The experiment is performed in such a way
that the driving frequency is fixed
and the temperature is changed to obtain the resonance condition.

Second and more seriously is the fact that the Higgs mode
is not the only process leading to a resonance in the THG spectrum.
Due to the driving, charge density fluctuations (CDF)
are excited as well
which resonate at the pair-breaking energy $2\Delta$
coinciding with the Higgs mode.
Depending on the system considered,
these contributions may exceed the Higgs contribution
by several orders \cite{PhysRevB.93.180507};
however, an analysis beyond BCS approximation shows
that the situation may reverse again \cite{PhysRevB.94.224519}.
Studying the polarization dependence of the THG signal
can help to distinguish between the contributions,
which is however not as clear as to be wished
as the polarization dependence depends strongly on the dispersion
and pairing interaction \cite{PhysRevB.97.094516}.
However, other papers pointed out that for superconductors in the dirty-limit
the paramagnetic coupling to light plays a crucial role
which leads to a dominant Higgs contribution
in the THG signal
\cite{JPhysSocJpn.84.114711,JPhysSocJpn.87.024704,PhysRevB.96.155311,%
PhysRevB.99.224510,PhysRevB.99.224511}.
This shows that a careful analysis of an experiment has to be performed
in order to interpret the data correctly.
In this work,
we concentrate on the Higgs contribution to THG response
and the influence of nontrivial pairing symmetry.
Nevertheless,
we calculate the charge-density fluctuations,
which dominate the total THG response within our clean-limit analysis.
As long as impurity scattering does not introduce any preferred direction,
our results about polarizations dependencies
should be valid even in the dirty-limit.

In this paper, we generalize the analysis of \cite{PhysRevB.92.064508}
within the Anderson pseudospin formulation of BCS theory
to arbitrary gap symmetry
and calculate the induced gap oscillations and the induced current.
As already shown in \cite{PhysRevB.95.104503} for a two-band superconductor,
the THG intensity contains resonances for each band
and for the relative phase oscillation -- the Leggett mode.
Thus, more complex systems can lead to additional features in the spectrum
mapping out the underlying structure.

We show, how an arbitrary gap symmetry affects the THG intensity
and that composite gap-symmetries can lead to additional resonances.
In addition we show how an asymmetric driving scheme
may resonate with additional asymmetric Higgs modes
leading to multiple resonances in the THG spectrum.
Such a driving scheme was inspired by recent experimental findings
\cite{arxiv.1901.06675}
and tries to model the asymmetry in driving due to small in-plane components
of the external field.

This paper is organized as follows.
In Sec.~\ref{sec:hamiltonian} we introduce the model
and explain the time-evolution via Bloch equations.
In Sec.~\ref{sec:gap_osci} we derive analytic expressions
for the driven Higgs oscillations and compare the $s$- and $d$-wave case.
In Sec.~\ref{sec:thg} we calculate analytic expressions
for the THG response as a function of the driving frequency.
In Sec.~\ref{sec:temp_dep} we calculate numerically
the temperature dependence of the THG response
for $s$-, $d$- and $d$+$s$-wave.
The following Sec.~\ref{sec:polarization}
evaluates the polarization dependence of the THG signal.
In Sec.~\ref{sec:asym_driving} we propose an asymmetric driving scheme
and show that an additional resonance peak in the THG signal occurs.
Finally we summarize and discuss the results in Sec.~\ref{sec:conclusion}.

\section{Hamiltonian and time evolution}
\label{sec:hamiltonian}
The starting point of our study
is the minimal model for Higgs oscillations,
i.e. the mean-field BCS-like Hamiltonian,
where we allow a $\vk$-dependent energy gap
\begin{align}
    H &= \sum_{\vk\sigma}
        \epsilon_\vk c^\dagger_{\vk\sigma}c_{\vk\sigma}
        \notag\\&\quad
        - \sum_\vk \left(
            \Delta_\vk c_{\vk\uparrow}^\dagger c_{-\vk\downarrow}^\dagger
            + \Delta_\vk^* c_{-\vk\downarrow}c_{\vk\uparrow}
        \right)\,.
\end{align}
Hereby, $\epsilon_\vk = \epsilon_{-\vk}$ is the energy dispersion
measured from the Fermi energy $\epsilon_{\mathrm F}$,
$c^\dagger_{\vk\sigma}$ and $c_{\vk\sigma}$
are the electron creation or annihilation operators and
\begin{align}
    \Delta_\vk = \sum_{\vk'} V_{\vk\vk'} \braket{
        c_{-\vk'\downarrow}c_{\vk'\uparrow}
    }
    = \Delta f_\vk
\end{align}
the momentum-dependent energy gap,
where we assume a separable pairing interaction
$V_{\vk\vk'} = V f_\vk f_{\vk'}$
with strength $V$ and the symmetry function $f_\vk$ such that
\begin{align}
    \Delta = V\sum_{\vk} f_{\vk}
        \braket{c_{-\vk\downarrow}c_{\vk\uparrow}}\,.
\end{align}
Hereby, we assume that $f_\vk$ is normalized to 1.
With this,
$\Delta$ corresponds to the maximum of the absolute value of the gap.
We implicitly restrict all summations over $\vk$
in the range $-\vk_c < \vk < \vk_c$
with the momentum cutoff $\vk_c$
and number of $\vk$-points $N$.
For the sake of simplicity we will restrict all of our calculations
to two dimensions having also the quasi-2d layered cuprates in mind.
Coupling to an electric field is incorporated via minimal substitution
$\epsilon_\vk \rightarrow \epsilon_{\vk-e\vec A(t)}$
with the electron charge $e$ and the vector potential
\begin{align}
    \vec A(t) = \VV{A_x(t)}{A_y(t)}
        = A_0(t) \hat e_A
        = A_0\sin(\Omega t) \VV{\cos\theta}{\sin\theta}
        \label{eq:At}
\end{align}
where $A_0$ is the driving amplitude,
$\Omega$ the driving frequency and $\theta$ the polarization angle.
The choice of our vector potential is such
that its wave vector is perpendicular to the superconducting plane
without an in-plane component.
This choice matches the experiments done so far
for the conventional $s$-wave superconductor NbN
\cite{Science.345.1145,PhysRevB.96.020505}.
However, small in-plane components could be induced
due to non-perfect alignment, tilted laser pulses
or higher-order nonlinear couplings leading to an asymmetric driving scheme.
This idea and its effects will be addressed in Sec.~\ref{sec:asym_driving}.

For our further analysis we will make use
of Anderson pseudospin formalism  \cite{PhysRev.112.1900}.
We introduce the Nambu-Gorkov spinor
$\Psi_\vk^\dagger = (c_{\vk\uparrow}^\dagger \,\,\, c_{-\vk\downarrow})$
and define Anderson pseudospin
\begin{align}
    \vec \sigma_{\vk} = \frac{1}{2} \Psi_{\vk}^\dagger \vec\tau \Psi_{\vk}
    = \frac 1 2 \VVV{c_{-\vk\downarrow}c_{\vk\uparrow}
    + c_{\vk\uparrow}^\dagger c_{-\vk\downarrow}^\dagger}
    {\im \left(c_{-\vk\downarrow} c_{\vk\uparrow}
    -  c_{\vk\uparrow}^\dagger c_{-\vk\downarrow}^\dagger\right)}
    {c_{\vk\uparrow}^\dagger c_{\vk\uparrow}
    - c_{-\vk\downarrow} c_{-\vk\downarrow}^\dagger}
\end{align}
with $\vec\tau$ the vector of Pauli matrices.
The BCS Hamiltonian can then be rewritten in terms of the pseudospin
and takes the form of a Hamiltonian
for a magnetic spin in an external magnetic field
\begin{align}
    H &= \sum_{\vk} \vec b_{\vk} \vec \sigma_{\vk}
\end{align}
with the definition of the pseudomagnetic field
\begin{align}
    \vec b_\vk &= \VVV{-2\Delta'f_\vk}
        {2\Delta''f_\vk}
        {\epsilon_{\vk-e\vec A(t)}+\epsilon_{\vk+e\vec A(t)}}
\end{align}
and $\Delta = \Delta' + \im\Delta''$.
The gap equation takes the form
\begin{align}
    \Delta &= V\sum_\vk f_\vk \Big(
        \braket{\sigma_\vk^x}
        - \im \braket{\sigma_\vk^y}
    \Big)\,.
    \label{eq:gap_eq_full}
\end{align}
In the equilibrium situation with $A_0 = 0$ at temperature $T$,
all the pseudospins are aligned parallel to the pseudomagnetic field.
Therefore, the equilibrium expectation values of the pseudospins read
\begin{align}
    \braket{\sigma_{\vk}^x}(0) &=
    \frac{\Delta f_\vk}{2E_{\vk}}
        \tanh\left(\frac{E_\vk}{2k_BT}\right)\,,\\
    \braket{\sigma_{\vk}^y}(0) &= 0\,,\\
    \braket{\sigma_{\vk}^z}(0) &=
    -\frac{\epsilon_{\vk}}{2E_{\vk}}
        \tanh\left(\frac{E_\vk}{2k_BT}\right)\,,\\
    \Delta(0) &= \Delta = V \sum_\vk f_\vk \braket{\sigma_\vk^x}(0)
\end{align}
with the quasiparticle energy
\begin{align}
    E_\vk = \sqrt{\epsilon_\vk^2 + (\Delta f_\vk)^2}\,.
\end{align}
Hereby, we assumed a real equilibrium gap,
i.e. $\Delta'' = 0$ and $\Delta = \Delta'$.
We use the following ansatz to describe the deviations
$x_\vk(t)$, $y_\vk(t)$, $z_\vk(t)$ and $\delta\Delta(t)$
from the equilibrium pseudospin expectation values and the gap
\begin{align}
    \braket{\sigma^x_\vk}(t) &= \braket{\sigma_\vk^x}(0) + x_\vk(t)\,,\\
    \braket{\sigma^y_\vk}(t) &= \braket{\sigma_\vk^y}(0) + y_\vk(t)\,,\\
    \braket{\sigma^z_\vk}(t) &= \braket{\sigma_\vk^z}(0) + z_\vk(t)\,,\\
    \Delta(t) &= \Delta + \delta\Delta(t)\,,\\
    \delta\Delta(t) &= V \sum_\vk f_\vk \Big(
        x_\vk(t) - \im y_\vk(t)
    \Big)\,.\label{eq:gap_eq_dD}
\end{align}
The time evolution of the expectation values
is governed by Heisenbergs' equation of motion
\begin{align}
    \partial_t \vec \sigma_{\vk}(t)
    = \im [H, \vec \sigma_\vk]
\end{align}
which take the form of Bloch equations
\begin{align}
    \VVV{\dot x_\vk(t)}{\dot y_\vk(t)}{\dot z_\vk(t)}
    = \vec b_{\vk}(t) \times \braket{\vec \sigma_{\vk}}(t)\,.
\end{align}
Note that we have set $\hbar = 1$ here and in the following.

For the further calculation,
we expand the $z$-component of the pseudomagnetic field
up to second order in $A_0$
\begin{align}
    \epsilon_{\vk - e\vec A(t)} + \epsilon_{\vk + e\vec A(t)}
        &= 2\epsilon_\vk + \epsilon_{\vk}^A(t) + \mathcal O(A_0^4)
\end{align}
with the definition
\begin{align}
    \epsilon_{\vk}^A(t)
        &= e^2 \sum_{ij} A_i(t) A_j(t)\partial^2_{ij} \epsilon_\vk
\end{align}
where we used the short notation
$\partial^2_{ij}\epsilon_\vk =
    \frac{\partial^2 \epsilon_\vk}{\partial_{k_i}\partial_{k_j}}$.
Due to the assumption $\epsilon_\vk = \epsilon_{-\vk}$,
the term linear in $A_0$ vanishes and only the quadratic coupling remains.
Written out explicitly, we find
\begin{align}
    \epsilon^A_\vk(t) &= e^2A^2_0(t)D_{\epsilon_\vk}(\theta)
    \intertext{with}
    D_{\epsilon_\vk}(\theta) &=
        \cos^2\theta \,\partial^2_{xx}\epsilon_\vk
        + \sin^2\theta \,\partial^2_{yy}\epsilon_\vk
        \notag\\&\qquad
        + 2\sin\theta\cos\theta \,\partial^2_{xy}\epsilon_\vk
    \,.\label{eq:Dek}
\end{align}
The term $D_{\epsilon_\vk}(\theta)$ contains the polarization dependence
and the second derivatives of the dispersion.
We can see that a non-parabolicity is crucial for the coupling
as a parabolic dispersion would lead to a $\vk$-independent coupling term
describing only a time-dependent variation of the Fermi energy.
This changes in the dirty-limit,
where also the paramagnetic coupling term remains
leading to a much stronger coupling
\cite{PhysRevB.99.224510,PhysRevB.99.224511}.
The explicit form of the Bloch equations used in the following reads
\begin{align}
    \braket{\vec{\dot\sigma}_{\vk}}(t)
    &= \VVV{-2(\Delta + \delta\Delta'(t))f_\vk}
        {2\delta\Delta''(t)f_\vk}
        {2\epsilon_\vk + \epsilon^A_\vk(t)}
    \times
    \VVV{\braket{\sigma_\vk^x}(0) + x_\vk(t)}
        {y_\vk(t)}
        {\braket{\sigma_\vk^z}(0) + z_\vk(t)}
    \label{eq:Bloch}\,.
\end{align}

\section{Higgs oscillations}
\label{sec:gap_osci}
To gain a first insight into the driven dynamics
and see the differences to the known $s$-wave case,
we start by deriving analytic expressions for the gap oscillations.
To this end, we assume a small driving amplitude $eA_0 \ll 1$,
such that we can neglect terms
in second order of the deviation from equilibrium.
For simpler notation,
we will show only the results for the zero temperature case $T = 0$.
A generalization for finite temperature is straight forward
by including the factor $\tanh(E_\vk/(2k_BT))$ in the respective expressions
and numerical results for this case are shown in the later sections.
The linearized equations of motion read
\begin{align}
    \dot x_\vk(t) &= -2\epsilon_\vk y_\vk(t)
        - \frac{f_\vk}{E_\vk}\epsilon_\vk\delta\Delta''(t)\,,\\
    \dot y_\vk(t) &= 2\epsilon_\vk x_\vk(t)
        + 2f_\vk\Delta z_\vk(t)
        \notag\\&\quad
        + \frac{f_\vk}{E_\vk}\left(
            \frac 1 2 \epsilon^A_\vk(t)\Delta
            - \epsilon_\vk\delta\Delta'(t)
        \right)\,,\\
    \dot z_\vk(t) &= -2f\Delta y_\vk(t)
        - \frac{\Delta f_\vk^2}{E_\vk}\delta\Delta''(t)\,.
\end{align}
We perform a Laplace transform
from time $t$ to complex frequencies $s$ according to
\begin{align}
    a(t) \rightarrow a(s) = \int_0^\infty \e^{st} a(t)\,\mathrm dt
\end{align}
to obtain algebraic equations
\begin{align}
    s x_\vk(s) &= -2\epsilon_\vk y_\vk(s)
        - \frac{f_\vk}{E_\vk}\epsilon_\vk\delta\Delta''(s)\,,\\
    s y_\vk(s) &= 2\epsilon_\vk x_\vk(s)
        + 2f_\vk\Delta z_\vk(s)
        \notag\\&
        + \frac{f_\vk}{E_\vk}\left(
            \frac 1 2 \epsilon^A_\vk(s)\Delta
            - \epsilon_\vk\delta\Delta'(s)
        \right)\,,\\
    s z_\vk(s) &= -2f\Delta y_\vk(s)
        - \frac{\Delta f_\vk^2}{E_\vk}\delta\Delta''(s)\,.
    \label{eq:bloch_linear_laplace}
\end{align}
Solving for the deviation terms, we find
\begin{align}
    x_\vk(s) &= \frac{
        \epsilon_\vk f_\vk\Big(
            2\epsilon_\vk\delta\Delta'(s)
            - \Delta\epsilon^A_\vk(s)
            - s\delta\Delta''(s)
        \Big)
        }{
            E_\vk (4E_\vk^2 + s^2)
        }\,,\label{eq:xk_sol}\\
    y_\vk(s) &= \frac{
        - f_\vk\Big(
            2s\epsilon_\vk\delta\Delta'(s)
            - s\Delta\epsilon^A_\vk(s)
            + 4E_\vk^2 \delta\Delta''(s)
        \Big)
        }{
            2 E_\vk (4E_\vk^2 + s^2)
        }\,,\label{eq:yk_sol}\\
    z_\vk(s) &= \frac{
        \Delta f_\vk^2\Big(
            2\epsilon_\vk\delta\Delta'(s)
            - \Delta\epsilon^A_\vk(s)
            - s\delta\Delta''(s)
        \Big)
        }{
            E_\vk (4E_\vk^2 + s^2)
        }\label{eq:zk_sol}\,.
\end{align}
To proceed with a further analytic analysis
we use several assumptions and approximations,
e.g. isotropy in $x$- and $y$-direction
(for details see appendix~\ref{sec:expansion_driving_term}).
With these assumptions, we can simplify the interaction term
$\epsilon_\vk^A(t)$ under a summation with an arbitrary factor $a_\vk$
which becomes polarization independent
\begin{align}
    \sum_\vk \epsilon_\vk^A(t) a_\vk
        &= \sum_\vk e^2A_0^2(t) (\alpha_0 + \alpha_1\epsilon_\vk) a_\vk\,.
\end{align}
Hereby, $\alpha_i$ are expansion coefficients.
To evaluate the momentum sum in the gap equation Eq.~\eqref{eq:gap_eq_full},
we use the approximation that the dispersion
depends only on the absolute value of $\vk$ near $\epsilon_{\mathrm F}$,
i.e. $\epsilon_\vk = \epsilon(|\vk|)$,
whereas the gap symmetry function only depends on the polar angle $\varphi$,
i.e $f_\vk = f(\varphi)$.
Then, we can replace the momentum sum
with an integral over the energy $\epsilon$
and an integral over the polar angle $\varphi$
\begin{align}
    V \sum_\vk \rightarrow
    \lambda \int_{-\epsilon_c}^{\epsilon_c} \,\mathrm d\epsilon \,
    \int_0^{2\pi} \,\mathrm d\varphi
\end{align}
with $\lambda = V D(\epsilon_{\mathrm{F}})$,
where the density of states $D(\epsilon_{\mathrm{F}})$
is assumed to be constant near the Fermi energy.
In the following we will replace the integral borders
of the $\epsilon$ integral with $\pm \infty$
which is valid for $\epsilon_c \gg \Delta$.
For further notation simplification,
we will drop the energy and angular dependencies
of the functions in the integrands.

Next, we insert the expressions for $x_\vk(t)$ and $y_\vk(t)$
into the real and imaginary part of the gap equation
\begin{align}
    \delta\Delta'(s) &= \lambda \int \mathrm d\epsilon
        \,\int\mathrm d\varphi \,
            f\,x(s) \label{eq:dD1}\,,\\
    \delta\Delta''(s) &= -\lambda \int \,\mathrm d\epsilon
        \,\int\,\mathrm d\varphi \,
            f\, y(s) \label{eq:dD2}\,.
\end{align}
We start by solving the equation for the real part of the gap.
The terms $\propto \delta\Delta''(s)$ and $\propto \alpha_0$
in Eq.~\eqref{eq:dD1} vanish as
\begin{align}
    \int \frac{\epsilon}{E(4E^2 + s^2)}\,\mathrm d\epsilon = 0
\end{align}
due to the asymmetry of the integrand. The resulting expression reads
\begin{align}
    &\delta\Delta'(s) =
    \frac 1 2 \alpha_1 \Delta e^2 A_0^2(s)
    \notag\\&
    \times \Bigg(1
        - \frac{1}{\lambda \int \,
        f^2(4\Delta^2 f^2 + s^2) F(s,\varphi)
        \,\mathrm d\varphi
        }
    \Bigg)
    \label{eq:dD1s}\,.
\end{align}
To obtain this result we identified the equilibrium gap equation
\begin{align}
    1 = \lambda \int \,\mathrm d\epsilon \int \,\mathrm d\varphi \,
        \frac{f^2}{2E}
\end{align}
by adding a zero $4\Delta^2 f^2 + s^2 - 4\Delta^2 f^2 - s^2$
in the nominator of the respective expression in the integrand.
The function $F(s,\varphi)$ is defined by
\begin{align}
    F(s,\varphi) &= \int \frac{1}{2E(4E^2 + s^2)}\,\mathrm d\epsilon\\
    &= \frac{1}{s\sqrt{4\Delta^2f^2 + s^2}}
    \sinh^{-1}\left(\frac{s}{2\Delta|f|}\right)\,.
\end{align}
Next, we use
\begin{align}
    A_0^2(s) = A_0^2\frac{2\Omega^2}{s(s^2+4\Omega^2)}
\end{align}
and define
\begin{align}
    I_1(s) &= \frac{\Omega^2}{s(s^2 + 4\Omega^2)}\,,\label{eq:I1I2_I1}\\
    I_2(s) &= \frac{\Omega^2}{(s^2+4\Omega^2)}
    \notag\\&\quad\times
    \frac{1}{\lambda\int
        f^2\sqrt{4\Delta^2f^2+s^2}
        \sinh^{-1}\left(\frac{s}{2\Delta|f|}\right)
        \,\mathrm d\varphi
    }
    \label{eq:I1I2_I2}
\end{align}
to write $\delta\Delta'(s) = \alpha_1 \Delta e^2 A_0^2 (I_1(s) + I_2(s))$.
The solution in the time domain
is obtained by performing the inverse Laplace transform.
The calculation can be found in the
appendix~\ref{sec:inverse_laplace_transform}.
The solution $\delta\Delta'(t)$ consists of three contributions
\begin{align}
    \delta\Delta'(t) &= \alpha_1 \Delta e^2 A_0^2 \Big(
        \delta\Delta_D(t) + \delta\Delta_H(t) + \delta\Delta_R(t)
    \Big)
    \label{eq:dD}
\end{align}
where
\begin{align}
    \delta\Delta_D(t) &= \frac{1-\cos(2\Omega t)}{4}\,,\\
    \delta\Delta_H(t) &= \frac{1}{\pi}\int_{-2\Delta}^{2\Delta}\mathrm dr\,
    \frac{\Omega^2}{4\Omega^2-r^2}
    \notag\\&\quad
    \times\frac{\sin(rt)}{\lambda \int\mathrm d\varphi\, f^2
        \sqrt{4\Delta^2 f^2 - r^2}
        \sin^{-1}\left(\frac{r}{2\Delta|f|}\right)
    }
    \notag\\&\quad
    - \frac{1}{4\lambda \int f^2\, \mathrm d\varphi}\,,\\
    \delta\Delta_R(t) &= \frac{\Omega\cos(2\Omega t)}{4\lambda
        \int\mathrm d\varphi\, f^2
        \sqrt{\Delta^2 f^2 -\Omega^2}
        \sin^{-1}\left(\frac{\Omega}{\Delta|f|}\right)
    }\label{eq:dDR}\,.
\end{align}
\begin{figure}[t]
    \includegraphics[width=0.5\textwidth]{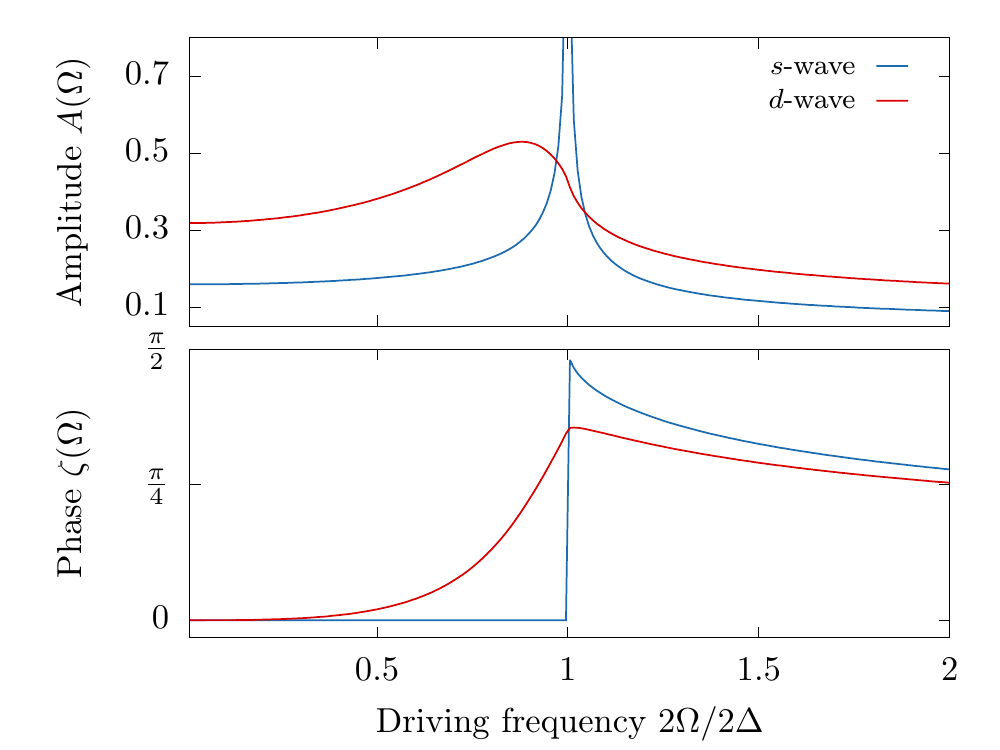}
    \caption{Amplitude $A$ and phase $\zeta$
    of the driven gap oscillation \eqref{eq:dDR2}.
    In the case of s-wave, a sharp resonance at $2\Omega = 2\Delta$ occurs,
    whereas for d-wave a broad peak slightly below $2\Delta$ is apparent.
    In addition, the phase of s-wave shows a sharp jump
    from $0$ for $2\Omega < 2\Delta$ to $\pi/2$ at $2\Omega = 2\Delta$
    and a drifting for $2\Omega > 2\Delta$.
    For d-wave, there is a broad phase change from $0$ for $\Omega = 0$
    up to a value slightly below $\pi/2$ for $2\Omega = 2\Delta$.
    At this point there is a sharp kink
    and a drifting for $2\Omega > 2\Delta$ similar to the s-wave case.
    }
    \label{fig:dD}
\end{figure}%
The term $\delta\Delta_D(t)$ is a forced oscillation of the gap
with twice the driving frequency $\Omega$
due to the nonlinear $A(t)^2$ driving.
The term $\delta\Delta_H(t)$ is the intrinsic Higgs oscillation
with a frequency of $\sim 2\Delta$
induced by an effective interaction quench
resulting from the periodic driving \cite{PhysRevB.92.064508}.
The amplitude of the Higgs oscillation
depends on the driving frequency $\Omega$.
Without solving the integral explicitely,
we can see that there will be a resonance when $\Omega = \Delta$
due to the prefactor $\Omega^2/(4\Omega^2 - r^2)$.
We can understand this as a coincidence of the two oscillations $2\Omega$
and $2\Delta$, when $\Omega$ is tuned to $\Delta$.
This resonance can be found also in the third term,
which is again a forced oscillation of $2\Omega$
where the amplitude is also frequency dependent.
We rewrite it as
\begin{align}
    \delta\Delta_R(t) &=
        \frac{1}{4\lambda} A(\Omega)\e^{-\im \zeta(\Omega)} \cos(2\Omega t)
    \label{eq:dDR2}
\end{align}
where $A(\Omega)$ represents the oscillation amplitude
and $\zeta(\Omega)$ the phase relative to the driving.

Let us first recapitulate the known result for the s-wave case
\cite{PhysRevB.92.064508},
where $f(\varphi) = 1$ and $\lambda \rightarrow \lambda/(2\pi)$.
In this case, there are two isolated branch points at $s = \pm 2\im\Delta$
which lead to an exact resonance in the amplitude if $2\Omega = 2\Delta$
(see appendix~\ref{sec:inverse_laplace_transform} for details),
i.e. the driving frequency resonates
with the intrinsic Higgs oscillation at $2\Delta$
\begin{align}
    A(\Omega) &= \frac{\Omega}{\sqrt{\Delta^2-\Omega^2}}
        \frac{1}{
            \sin^{-1}\left(\frac{\Omega}{\Delta}\right)
        }\,.
\end{align}
The amplitude $A(\Omega)$ and phase $\zeta(\Omega)$ of the resonance term
is plotted in Fig.~\ref{fig:dD}.
One can observe a sharp resonance
and also a sharp $\frac\pi 2$ phase jump at the resonance condition.

Next, let us consider the d-wave case where $f(\varphi) = \cos(2\varphi)$.
Here, we do no longer have an exact resonance condition
but there is still a maximum of the amplitude
given by the minimum of the integral in the denominator of \eqref{eq:dDR}.
This leads to a broad peak in the amplitude
accompanied by a smooth phase change of $\zeta(0)-\zeta(2\Delta) < \pi/2$.
The peak in the amplitude is at an energy slightly below $2\Omega=2\Delta$,
whereas are sharp kink in the phase at $2\Omega=2\Delta$ is found.

For nontrivial gap symmetry,
the gap oscillation spectrum does not have to be peaked at exactly $2\Delta$.
It depends on several factors
including the density of states
or the exact shape of the gap symmetry function.
In particular, here, the position of the peak is determined by the interplay
between the terms $f^2$ and $\sqrt{\Delta^2f^2-\Omega^2}$
under the integral over $\varphi$ in \eqref{eq:dDR}.
For values $\Omega \approx \Delta$,
the terms with the highest weights, i.e. $f \approx 1$,
vanish which decreases the value of the integral and creates its minimum.
Therefore, in the case of $d$-wave, the maximum of the amplitude
can be still found roughly at $2\Omega \approx 2\Delta$.
Depending on the exact shape of $f$, it can be shifted to lower energies.
For values $2\Omega > 2\Delta$, the square root is always imaginary
for each value of $\varphi$,
which leads to the sharp edge in the phase at $2\Omega = 2\Delta$.

After studying the real part,
we will finally also calculate the imaginary part
of the gap with Eq.~\eqref{eq:dD2}.
Similar to the real part,
the terms $\propto \delta\Delta'(s)$ and $\propto \alpha_1$ vanish
and we are left with
\begin{align}
    \delta\Delta''(s) &= -\alpha_0 \Delta e^2 A_0^2(s) s \lambda
        \int \mathrm d\varphi\, f^2 F(s,\varphi)
        \notag\\&
        + \delta\Delta''(s)
        - \delta\Delta''(s) s^2 \lambda
            \int \mathrm d\varphi\, f^2 F(s,\varphi)\,.
\end{align}
Solving for $\delta\Delta''(s)$ yields
\begin{align}
    \delta\Delta''(s) &= -\alpha_0 \Delta e^2 A_0^2
        \frac{2\Omega}{s^2(s^2+4\Omega^2)}
    \label{eq:dD2s}
\end{align}
and the solution in time-domain is given by
\begin{align}
    \delta\Delta''(t) &= -\alpha_0 \Delta e^2 A_0^2
        \left(\frac t 2 - \frac{\sin(2\Omega t)}{4\Omega}\right)
        \\
        &= -\alpha_0 \Delta e^2 \int_0^t A_0(t')^2 \,\mathrm dt'\,.
\end{align}
We can see that the imaginary part is independent of the gap symmetry.
It follows the oscillation of the driving frequency
with a frequency independent amplitude.
In addition it contains a drift linear in time.
Like in the $s$-wave case, for small driving amplitudes,
the imaginary part only contributes to the phase $\varphi_\Delta$
of the gap $\Delta(t) = |\Delta(t)|\e^{\im\varphi_{\Delta}(t)}$
\begin{align}
    \varphi_{\Delta}(t) &= \arctan\left(
        \frac{\delta\Delta''(t)}{\Delta + \delta\Delta'(t)}\right)
    \approx \frac{\delta\Delta''(t)}{\Delta}\,.
\end{align}
The absolute value of the gap is independent of the imaginary part
and is solely determined by the real part
\begin{align}
    |\Delta(t)| = \sqrt{(\Delta+\delta\Delta'(t))^2 + \delta\Delta''(t)^2}
    \approx \Delta+\delta\Delta'(t)\,.
\end{align}

\section{Third harmonic generation}
\label{sec:thg}
The resonance of the forced gap oscillation with the Higgs mode
can be found experimentally in the transmitted light field.
The nonlinear coupling of the vector potential to the condensate
leads to higher-harmonic generation
where the lowest non-vanishing order is third-harmonic generation.
Its intensity $I^{\mathrm{THG}}$,
which is proportional to the amplitude squared
of the induced current $j^{(3)}(3\Omega)$
\begin{align}
    I^{\mathrm{THG}} \propto \Big|j^{(3)}(3\Omega)\Big|^2
\end{align}
shows the same resonance as the gap oscillation
and will be calculated in the following.
However, not only the driven gap oscillations contribute to THG
but also charge density fluctuations,
i.e. single particle excitations
resonating at the pair-breaking energy $2\Delta$.
In the clean-limit, depending on material parameters,
these can be significantly larger in an experiment
and overlay the contribution from the gap oscillation.
Therefore, this contribution is considered
and calculated in this section as well
and compared to the contribution from the Higgs oscillation.

Driving the superconductor periodically will induce an electric current
\begin{align}
    \vec j(t) = e\sum_\vk \vec v_{\vk-e\vec A(t)} \braket{n_\vk}(t)
\end{align}
with the group velocity $\vec v_\vk = \nabla \epsilon_\vk$
and the charge density
\begin{align}
    \braket{n_\vk}(t) &= \braket{
        c_{\vk\uparrow}^\dagger c_{\vk\uparrow}
        + c_{\vk\downarrow}^\dagger c_{\vk\downarrow}
    }(t)\,.
\end{align}
To calculate the lowest order response,
we expand the velocity in $A_0$
\begin{align}
    v_{\vk-e\vec A(t)} &= \vec v_{\vk}
        - e \sum_j A_j(t) \partial_j \vec v_\vk + \mathcal O(A_0^2)\,.
\end{align}
The charge density can be expressed with the $z$-component of the pseudospin
and we obtain for the current
\begin{align}
    \vec j(t) = \vec j^{(0)}(t) + \vec j^{(1)}(t) + \vec j^{(3)}(t)
\end{align}
where
\begin{align}
    \vec j^{(0)}(t) & = e \sum_\vk \vec v_\vk\braket{n_\vk}(t)\,,\\
    \vec j^{(1)}(t) & = - 2e^2
        \sum_\vk \sum_j A_j(t) \partial_j \vec v_\vk
            \left(\braket{\sigma_\vk^z}(0) + \frac 1 2\right)\,,\\
    \vec j^{(3)}(t) & = - 2e^2 \sum_\vk \sum_j
        A_j(t) \partial_j \vec v_\vk z_\vk(t)\,.
\end{align}
The first term $\vec j^{(0)}(t)$ vanishes due to parity,
the second term $\vec j^{(1)}(t)$ represents the induced current
oscillating with the driving frequency $\Omega$.
The third term $\vec j^{(3)}(t)$ is the lowest order
of higher-order generation,
which oscillates with $3\Omega$
due to the proportionality $\propto A_j(t)z_\vk(t)$
as we have seen in the previous section
that $z_\vk(t)$ oscillates with $2\Omega$.
The induced current for an arbitrary angle
relative to the polarization of the vector potential
can be decomposed into a parallel and perpendicular component
which we will calculate separately.
We insert the expression for the vector potential
and expand the summation of the components
\begin{align}
    j^{(3)}_{\parallel\perp}(t)
        &= \vec j^{(3)}(t) \cdot \hat e_A^{\parallel\perp}\\
        &=- 2e^2 A_0(t) \sum_\vk
            D_{\epsilon_\vk}^{\parallel\perp}(\theta)z_\vk(t)
    \label{eq:j_parallel}
\end{align}
with $\hat e_A^\parallel = \hat e_A$,
$\hat e_A^\perp = (\sin\theta,-\cos\theta)^\top$,
$D_{\epsilon_\vk}^\parallel = D_{\epsilon_\vk}$ from Eq.~\eqref{eq:Dek} and
\begin{align}
    D_{\epsilon_\vk}^\perp &= \sin\theta\cos\theta
        (\partial_{xx}^2 \epsilon_\vk - \partial_{yy}^2 \epsilon_\vk)
        \notag\\&\quad
        + \partial_{xy}^2\epsilon_\vk(\sin^2\theta - \cos^2\theta)\,.
\end{align}
To obtain an expression for the spectrum, we perform a Fourier transform
and make use of the solution $z_\vk(s=\im\omega)$
from the linearized Bloch equations \eqref{eq:zk_sol}.
There are three different contributions
originating from the terms $\propto \delta\Delta'(s)$,
$\propto \delta\Delta''(s)$ and $\propto \epsilon_\vk^A(s)$
such that we can write
\begin{align}
    j^{(3)}_{\parallel\perp}(3\Omega) &=
        j^{(3)\text{H}}_{\parallel\perp}(3\Omega)
        + j^{(3)\text{P}}_{\parallel\perp}(3\Omega)
        + j^{(3)\text{CDF}}_{\parallel\perp}(3\Omega)\,,
\end{align}
where
\begin{align}
    j_{\parallel\perp}^{(3)\mathrm{H}}(3\Omega) &\propto \Delta
        A_0 \delta\Delta'(2\im\Omega)
        \label{eq:jHiggs}\,,\\
    j_{\parallel\perp}^{(3)\mathrm{P}}(3\Omega) &\propto \Delta
    A_0 \delta\Delta''(2\im\Omega)
        \label{eq:jPhase}\,,\\
    j_{\parallel\perp}^{(3)\mathrm{CDF}}(3\Omega)
        &\propto \Delta^2 A_0 A_0^2(2\im\Omega)\label{eq:jCDF}\,.
\end{align}
Details of the calculation and exact expression
can be found in appendix~\ref{sec:deriv_thg}.
The first term \eqref{eq:jHiggs} is the Higgs contribution following
from the amplitude oscillation of the order parameter.
The second term \eqref{eq:jPhase} follows from the oscillation of the
imaginary part of the gap, i.e. oscillations of the phase.
The third term \eqref{eq:jCDF} depends on the coupling of the
vector potential to the band dispersion and describes
charge density fluctuations as the expression has the form
of a density-density correlation function \cite{PhysRevB.93.180507}.

Using the same assumptions from the previous section
and rewriting occurring sums as integrals one obtains
\begin{figure}[t]
    \includegraphics[width=0.5\textwidth]{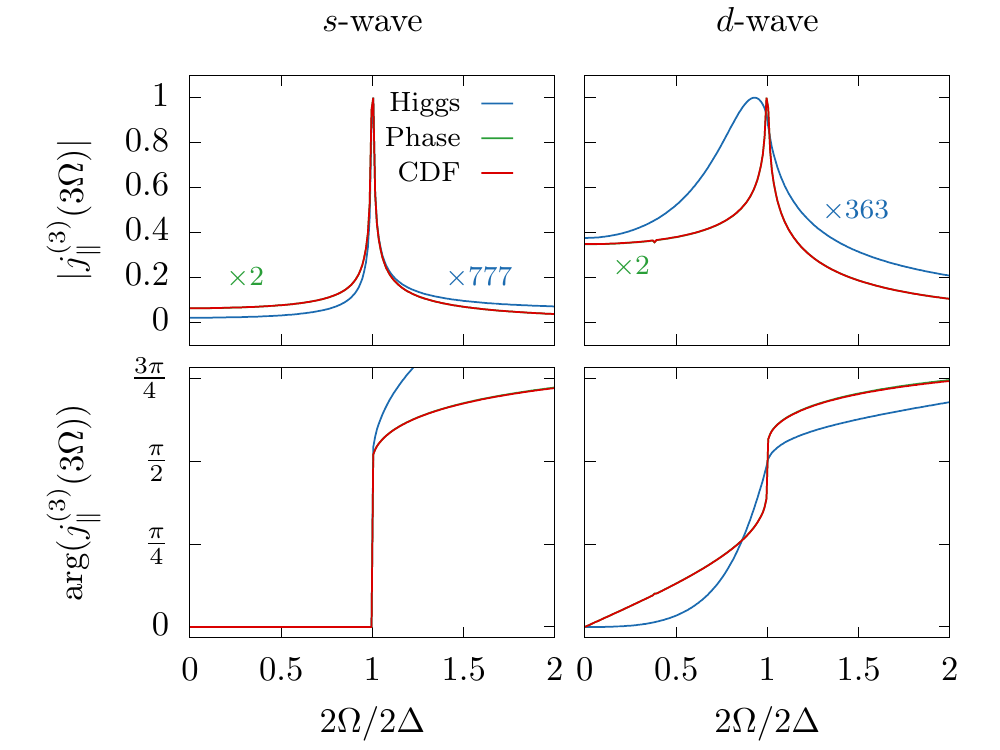}
    \caption{Individual contributions of the current
    $j^{(3)}_\parallel(3\Omega)$ for $s$- and $d$-wave
    originating from Higgs, CDF and phase oscillations
    using Eqs.~\eqref{eq:jHiggsApprox}, \eqref{eq:jPhaseApprox} and
    \eqref{eq:jCDFApprox}.
    The phase and CDF contribution lie approximately on top of each other.
    The parameters are $\Delta = 20$\,meV, $\theta = 0$, $\lambda = 0.1$ and
    $\alpha_0 = -\epsilon_{\mathrm F}/2$, $\alpha_1 = -0.5$
    with $\epsilon_{\mathrm F} = -400$\,meV
    assuming the tight-binding dispersion \eqref{eq:sbnn_dispersion}.
    }
    \label{fig:thg_omega}
\end{figure}%
\begin{align}
    j_{\parallel}^{(3)\mathrm H}(3\Omega) &\propto
        \alpha_1^2 \Delta^2 e^2 A_0^2(2\im\Omega)
        \notag\\&\quad
        \times \left(
            1
            - \frac{H(2\im\Omega)}{2}
            - \frac{1}{2H(2\im\Omega)}
        \right) \label{eq:jHiggsApprox}\,,\\
    j_{\parallel}^{(3)\mathrm P}(3\Omega) &\propto
        \alpha_0^2\Delta^2 e^2A_0^2(2\im\Omega) 2G(2\im\Omega)
        \label{eq:jPhaseApprox}\,,\\
    j_{\perp}^{(3)\mathrm{CDF}}(3\Omega) &\propto
        -\Delta^2 e^2 A_0^2(2\im\Omega)
        \frac 1 4 \sin 4\theta
        \notag\\&\quad
        \times \left(
            4\alpha_0^2 G(2\im\Omega)
            +\alpha_1^2(1-H(2\im\Omega))
        \right)\,,\\
    j_{\parallel}^{(3)\mathrm{CDF}}(3\Omega) &\propto
        -\Delta^2 e^2 A_0^2(2\im\Omega)
        \left(1-\frac 1 2 \sin^2 2\theta\right)
        \notag\\&\quad
        \times \left(
            4\alpha_0^2 G(2\im\Omega)
            +\alpha_1^2(1-H(2\im\Omega))
        \right) \label{eq:jCDFApprox}\,.
\end{align}
For these expressions we defined the integrals
\begin{align}
    H(2\im\Omega) &= \lambda \int \,
        f^2(4\Delta^2 f^2 - 4\Omega^2) F(2\im\Omega,\varphi)
        \,\mathrm d\varphi\,,\\
    G(2\im\Omega) &= \lambda \int \,
        f^2 F(2\im\Omega,\varphi)
        \,\mathrm d\varphi\,.
\end{align}
Each of the three terms for the current contribute to the THG intensity
\begin{align}
    I^{\mathrm{THG}}_{\parallel\perp} \propto
        \Big|j^{(3)}_{\parallel\perp}(3\Omega)\Big|^2\,.
\end{align}
Within the chosen assumptions we get an important result, namely that
the Higgs and phase contributions are polarization independent
and their contributions perpendicular to the polarization vector vanish.
In case of the CDF, both parallel and perpendicular contributions exist
and are polarization dependent with a characteristic dependency
independent of the gap symmetry and dispersion.
Both expressions $G(2\im\Omega)$ occurring in the phase and CDF terms
as well as $1/H(2\im\Omega)$ in the Higgs term diverge for $2\Omega = 2\Delta$
in the $s$-wave case or show a maximum in the $d$-wave case.
Therefore, the resonance of the amplitude oscillation with the Higgs mode
is not the only cause of the peak found in the THG.
In particular, the individual strengths are determined
by the parameters $\lambda$ and $\alpha_0$.

For $\alpha_0 = 0$, e.g. the half-filling case in the tight-binding model,
the resonance due to the $G$ term vanishes,
which removes the phase contribution completely
and strongly suppresses the CDF term.
For a polarization value of $\theta = \pi/4$,
the CDF contribution perpendicular vanishes
and the diverging part due to $G$ in the parallel contribution
exactly cancels with the phase contribution.
The polarization dependence will be discussed
in Sec.~\ref{sec:polarization} in more detail.
For large $\lambda$, i.e. large interaction strength,
the CDF contribution is enhanced over the Higgs contribution.
It can be understood in this way
that the resonance term for the Higgs contribution
scales with $\frac 1 \lambda$ due to the $\frac 1 H$ term,
whereas the resonance term for the CDF contribution
scales with $\lambda$ due to the $G$ term.
Therefore one can roughly estimate that
$j_\parallel^{(3)H}(3\Omega) \propto 1/\lambda^2 j_\parallel^{(3)CDF}(3\Omega)$.

In Fig~\ref{fig:thg_omega}, a comparison of the individual contributions
for the $s$- and $d$-wave case is shown
by evaluating the expressions \eqref{eq:jHiggsApprox},
\eqref{eq:jPhaseApprox} and \eqref{eq:jCDFApprox} for typical parameters.
We can see that the CDF term exceeds the Higgs term
by more than two orders of magnitude.
Within the chosen approximations and parameters,
the phase term is roughly $1/2$ of the CDF term.
Both terms show a sharp peak at $2\Omega=2\Delta$
for the $s$- and $d$-wave case.
This results from the resonance in the $G(2\im\Omega)$
at the pair-breaking energy $2\Delta$ even for the $d$-wave case
due to the much higher weight at the antinodes relative to the nodes.
The shape of the Higgs term,
originating from the resonance
in the amplitude oscillation $1/H(2\im\Omega)$,
follows the shape shown in Fig.~\ref{fig:dD},
i.e. a sharp peak for $s$-wave and a broad peak for $d$-wave.
Again for $d$-wave, there is also a lot of weight
in the range $2\Omega < 2\Delta$
as for any $\Omega$ there is always a $\Delta_\vk$
where $2\Omega = 2\Delta_\vk$ leading to an enhancement of the amplitude.
The phase change in the $s$-wave case is very sharp for all contributions,
while for the $d$-wave case there are small differences.
The phase change in the Higgs term
is similar to the phase change of the amplitude oscillation,
while the phase change of the CDF and phase term
is smooth in the beginning but contains a steep step around
$2\Omega = 2\Delta$.

Despite the fact that the CDF contribution
may exceed the Higgs contribution in our simple analysis,
it is still useful for the understanding of the physical mechanisms.
The actual weighting of the terms in an experiment
depends strongly on the material by further effects
not considered in our analysis,
like retardation effects in materials with phonon-mediated interaction
\cite{PhysRevB.94.224519}
or the paramagnetic coupling for superconductors in the dirty-limit
\cite{JPhysSocJpn.84.114711,JPhysSocJpn.87.024704,PhysRevB.96.155311,%
PhysRevB.99.224510,PhysRevB.99.224511}.

After gaining a first understanding of the terms
contributing to the THG intensity under the chosen assumptions
for $\epsilon_\vk$ and $f_\vk$,
we will drop these in the following sections
and solve the summations numerically without approximation
and include the temperature dependence.
This allows for arbitrary dispersion and gap symmetries to be considered
which can introduce new features
like additional resonances and polarization dependencies.

\section{Temperature dependence}
\label{sec:temp_dep}
In all current experiments so far, the driving frequency
cannot be tuned continuously like done in the theoretical analysis
from the previous section.
In order to find the resonance, the driving frequency is fixed
and the temperature is varied until $2\Omega = 2\Delta(T)$ is fulfilled
\cite{Science.345.1145}.
For the following, we solve the sums
\eqref{eq:xsum0}-\eqref{eq:xsum6} numerically without further approximations.
The interaction strength $V$ is calculated
for a chosen initial energy gap at $T=0$.
Then, the temperature dependence of the energy gap $\Delta(T)$
is determined self-consistently for each temperature.
To handle the divergences in the summations
and keep the required momentum grid resolution in a reasonable range,
we introduced a residual broadening of
$2\im\Omega \rightarrow 2\im\Omega + 0.01\Delta$.
This slightly broadens the resonance peaks
and washes out the sharp phase jumps but does not change the overall result.
We discretize the momentum space around the Fermi energy
with an energy cutoff of $E_c = 2\Delta$
with $N_k=2000$ points in $k$
and $N_\varphi = 2000$ points in the angular direction.
For the following calculation we use a $d$-wave gap function
$\Delta(\varphi) = \Delta \cos(2\varphi)$ with $\Delta = 20$\,meV
and the tight-binding dispersion \eqref{eq:sbnn_dispersion}
with $t = 200$\,meV and $\epsilon_{\mathrm F} = -400$\,meV.

\begin{figure}[t]
    \includegraphics[width=0.5\textwidth]{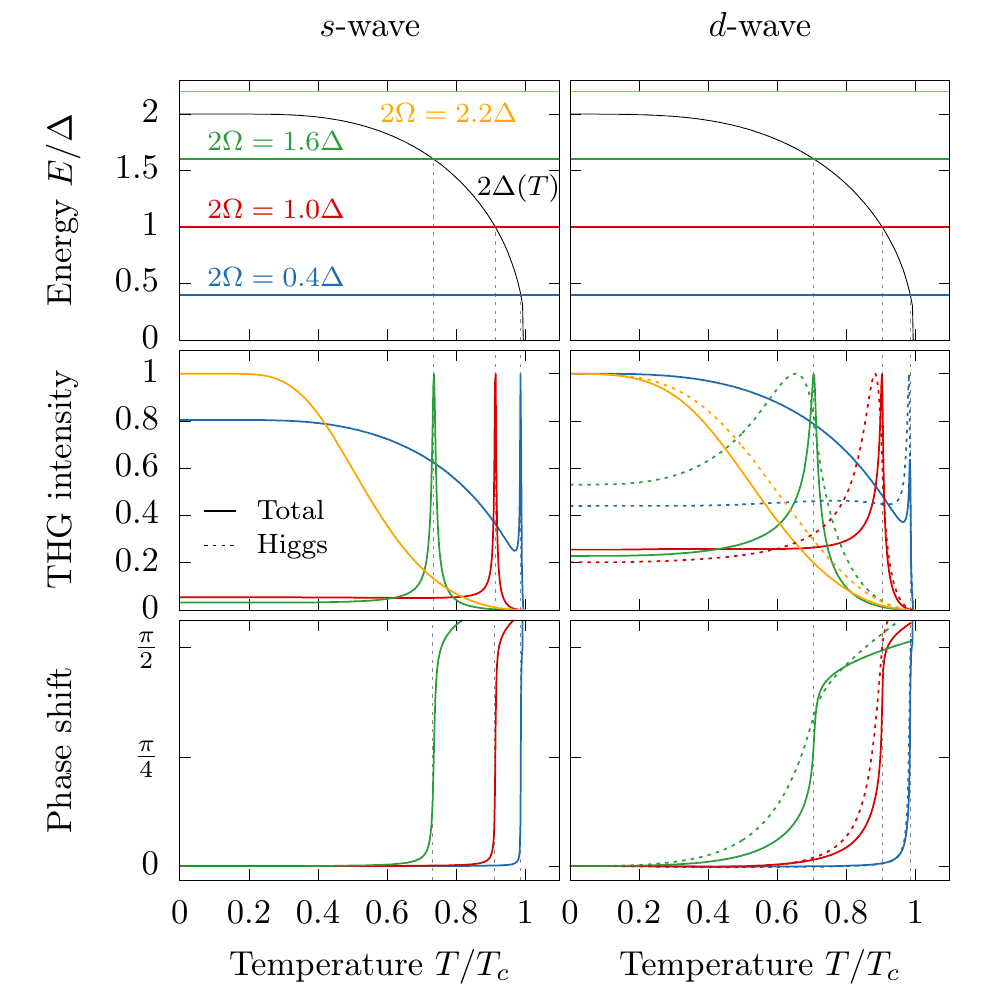}
    \caption{Temperature dependence of THG intensity
    for $s$- and $d$-wave symmetry for four different driving frequencies.
    The upper row shows the driving frequencies
    in relation to the energy gap.
    The middle row shows the THG intensity
    and the bottom row the phase.
    The polarization is $\theta = 0$.
    The solid lines are the full THG intensity,
    the dotted lines only the Higgs contribution.
    All intensities are normalized to their individual maximum.}
    \label{fig:thg_temp}
\end{figure}

In Fig.~\ref{fig:thg_temp}, the temperature dependence
of the THG intensity and phase for $d$-wave
in comparison with $s$-wave is shown for different driving frequencies.
For driving frequencies $\Omega > \Delta(T = 0)$
no resonance occurs and the THG intensity
follows roughly the temperature dependence
of the energy gap to the power of four, i.e.
$\sim\Delta(T)^4$.
As soon as $\Omega < \Delta(T=0)$, there is a temperature,
where $\Omega = \Delta(T)$ and a resonance occurs.
There are two main differences in the intensities of $s$- and $d$-wave
for the Higgs contribution (dotted lines).
First, for the same driving frequency
the resonance peaks are broader for $d$-wave
as there is no single resonance point like in the $s$-wave case.
Second, due to the continuous variation of the gap
from $0$ to $\Delta$ for $d$-wave,
there is always some resonant gap for temperatures $T < T_{\mathrm R}$,
where $T_{\mathrm R}$ is the temperature
for which $\Omega = \Delta(T_{\mathrm R})$.
This leads to a larger THG intensity background in that temperature range.
One can also see that for the chosen parameters
the CDF contribution dominates
as there is still a sharp resonance peak in the total $d$-wave intensity.

Even though the calculation for a pure $s$-wave order parameter
compared to a pure $d$-wave order parameter shows a clear
difference in the peak to background ratio for the Higgs contribution,
a comparison of intensities in an experiments is difficult.
On the one hand,
there is nothing to compare to for a given material.
On the other hand, peak intensities can vary due to additional damping effects,
which makes a clear distinguishing very unreliable.

The phase analysis confirms the analytic result in Fig.~\ref{fig:thg_omega}.
For $s$-wave a sharp phase jump of $\pi/2$ occurs at the resonance,
whereas for $d$-wave the phase jump is much broader.

As an example for what new features a more exotic gap symmetry may introduce,
we will study a case with $d$+$s$-symmetry,
i.e. $d$-wave with an admixture of $s$-wave
as it was observed for overdoped YBCO
\cite{NatPhys.2.190,PhysRevB.80.064505,PhysRevB.75.174525,%
PhilosMagLett.82.279}.
Following \cite{PhysRevB.75.174525}, we choose the gap
\begin{align}
    \Delta f(\varphi) = \Delta_d\cos(2\varphi) + \Delta_s
\end{align}
with $\Delta_d = 0.9\Delta$ and $\Delta_s=0.1\Delta$.
\begin{figure}[t]
    \includegraphics[width=0.5\textwidth]{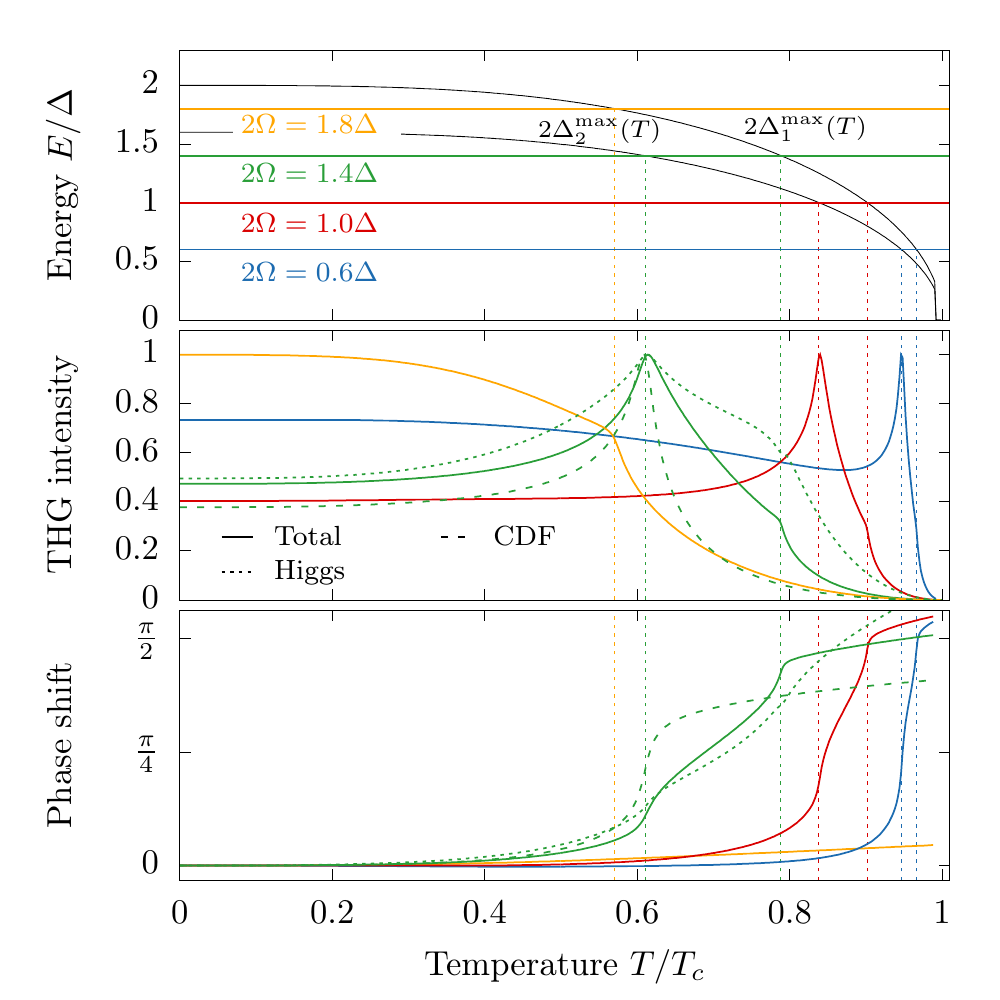}
    \caption{Temperature dependence of THG intensity for $d$+$s$-symmetry.
    The upper row shows the driving frequency
    in relation to the temperature dependent maxima of the energy gap.
    The second row shows the total THG intensity with solid lines
    and exemplary for one driving frequency
    the respective CDF (dotted) and Higgs (dashed) contribution.
    The intensity values are individually normalized to their maximum.
    The third row shows the phase change.
    }
    \label{fig:thg_dps}
\end{figure}%

The temperature dependence of the THG intensity
is shown in Fig.~\ref{fig:thg_dps}.
As the absolute value of the $d$+$s$ gap contains two local gap maxima,
i.e. $\Delta_1^{\mathrm{max}}=\Delta_d + \Delta_s = \Delta$
and $\Delta_2^{\mathrm{max}}=\Delta_d-\Delta_s = 0.8\Delta$,
we show the temperature dependence of these two curves.
We can see that if $\Omega < \Delta_2^{\mathrm{max}}$,
two resonances occur when the driving frequency matches these two maxima.
This can be seen both in the THG intensity as two peaks
as well as in the phase,
where a broad phase transition occurs
over the range of the two resonance temperatures
with sharp kinks at the resonance points.
\begin{figure}[t]
    \includegraphics[width=0.5\textwidth]{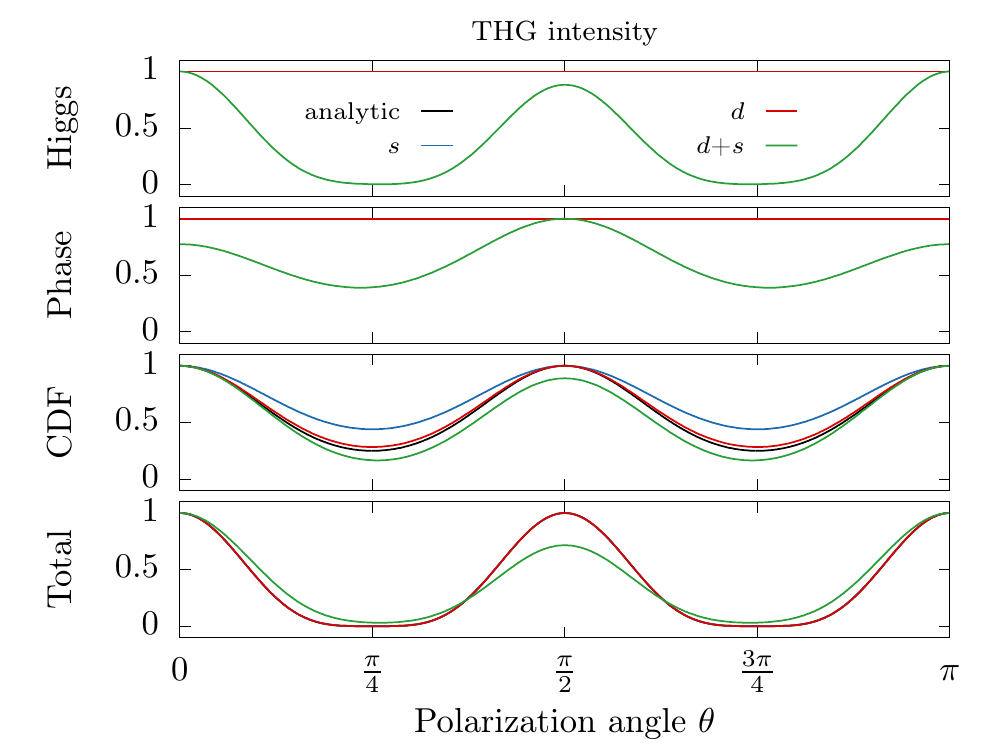}
    \caption{Polarization dependence of the Higgs contribution (first row),
    phase contribution (second row), CDF contribution (third row)
    and total THG intensity (fourth row).
    The analytic formulas
    from Eqs.~\eqref{eq:jHiggsApprox}-\eqref{eq:jCDFApprox},
    derived using the assumption of equivalence of $x$- and $y$-direction
    in $f_\vk^2$ and $\epsilon_\vk$,
    are compared with the numerically calculated results
    for $s$- and $d$-wave using dispersion Eq.~\eqref{eq:sbnn_dispersion}
    and $d$+$s$-wave using dispersion Eq.~\eqref{eq:disp_tb2_distorted}
    without approximation.
    All expressions are normalized to their maximum value.}
    \label{fig:thg_theta}
\end{figure}%

It is interesting to note that the two-peak structure
is an effect originating alone from the Higgs contribution
despite its smaller value.
While the CDF contribution only shows a single peak,
it is the Higgs contribution which shows the two-peak structure.
We can understand this as two Higgs modes at energies
$\Delta_1^{\mathrm{max}}$ and $\Delta_2^{\mathrm{max}}$
for each local gap maxima which resonate with the driving frequency.
This shows that even if the CDF contribution dominates,
the Higgs contribution may still contribute to specific features
visible in the spectrum.
If the polarization is tuned to $\theta = \pi/4$,
the CDF contribution contains a two peak structure as well
as the $G$ term in Eq.~\eqref{eq:jCDFApprox} is suppressed
due to the equivalent term with opposite sign in Eq.\eqref{eq:jPhaseApprox}
and the smaller $H$ term with the two-peak structure becomes visible.
We can conclude this section by stating that composite gap symmetries
can show additional resonances
if there are multiple local gap maxima with different amplitudes.

\section{Polarization dependence}
\label{sec:polarization}
As stated above, one possibility in an experiment
to gain more insight about the relative weight of Higgs and CDF contributions
is the polarization dependence.
As one can see in Eq.~\eqref{eq:jCDFApprox},
the CDF contribution has a very characteristic polarization dependence,
whereas the Higgs and phase terms
do not dependent on the polarization independent on the gap symmetry.
If there is no polarization dependence in an experiment
it can be a hint that the Higgs contribution is stronger than the CDF part.
This was observed in \cite{PhysRevB.96.020505}
where it was concluded that in NbN
the Higgs contribution dominates the THG intensity.
If we look at the expressions
\eqref{eq:jCDFApprox} and \eqref{eq:jPhaseApprox}
for the CDF and phase contribution of the current,
we can see that for $\theta = \pi/4$
the $G(2\im\Omega)$ term in the CDF expression cancels
exactly with the $G(2\im\Omega)$ term in the phase expression,
which means that for this particular polarization angle
only the Higgs contribution remains.

However, the derived formulas for the current
are only valid under the chosen special assumptions
$f_\vk^2$ and $\epsilon_\vk$ are symmetric under the exchange of
$k_x \leftrightarrow k_y$, as it is the case
for $s$- or $d$-wave symmetry and a tight-binding dispersion.
For the $d$+$s$ case this is no longer valid.
In Fig.~\ref{fig:thg_theta} the polarization dependence
for $s$-, $d$- and $d$+$s$-symmetry calculated numerically
is shown in comparison with the analytic result.
To show the influence of the dispersion,
we use a slightly distorted square lattice dispersion in the $d$+$s$-wave case
\begin{align}
    \epsilon_\vk &= -2t((1+\delta_0)\cos k_x + (1-\delta_0)\cos k_y)
        \notag\\&\qquad
        -4t'\cos k_x \cos k_y -\epsilon_{\mathrm F}
        \label{eq:disp_tb2_distorted}
\end{align}
with $\delta_0 = -0.03$ and $t=200$\,meV, $t'=-80$\,meV
and $\epsilon_{\mathrm F} = -240$\,meV \cite{PhysRevB.75.174525}.
One can see that for $s$- and $d$-wave
the result closely follows the approximation,
i.e. the Higgs contribution is polarization independent
and the CDF contribution has a
$\propto 1 - \frac 1 2 \sin^2 2\theta$ dependency
with a small offset resulting from the chosen approximations.
This changes for $d$+$s$,
where the Higgs and phase terms also becomes polarization dependent
and the polarization dependence of the total THG intensity
deviates from the analytic formula.
Choosing a different dispersion can lead to another polarization dependence.
We can see that the polarization dependence alone is not enough
to unambiguously distinguish between Higgs and CDF contributions
if the band dispersion is not exactly known.
This was also concluded in \cite{PhysRevB.97.094516},
where it was shown that the Higgs contribution
can get polarization dependent as well
while on the other hand the
polarization dependence of the CDF contribution can be suppressed.
Additional deviations from the derived formulas
for the polarization dependence are possible,
e.g. due to interplay with the pseudogap phase or multiband models.
This is however beyond the scope of this paper.

\section{Asymmetric driving}
\label{sec:asym_driving}
\begin{figure}[t]
    \includegraphics[width=0.5\textwidth]{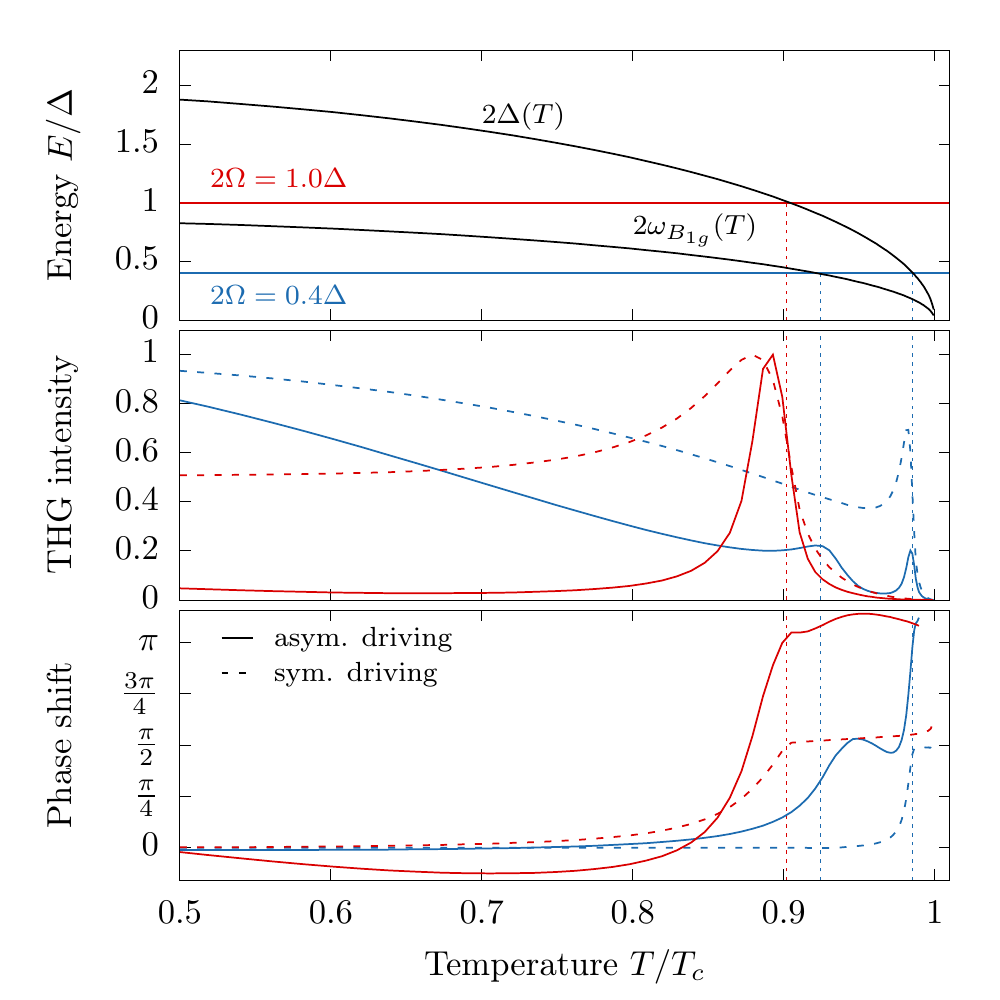}
    \caption{Temperature dependence of THG
    for $d$-wave for two different driving frequencies
    using the asymmetric driving scheme \eqref{eq:bk_asym}
    with $\delta_A=0.1$.
    The top row shows the driving frequencies
    in relation to the energy gap and the second mode.
    The middle row shows the THG intensity
    and the bottom row the phase.
    The result from the asymmetric driving (solid)
    is compared with the standard driving (dashed).}
    \label{fig:thg_asym}
\end{figure}%
In an experiment, the coupling of light to the superconducting condensate
may contain more subtleties and consists not only
of a symmetric diamagnetic $A^2$ term.
For examples, small in-plane components of the wavevector
induced by non perfect perpendicular alignment of laser and crystal
or on purpose tilted lasers
as well as higher order couplings to other finite-momentum modes
may induce an asymmetry while driving.
Such a momentum dependent driving can lead to asymmetric oscillations
of the condensate with respect to the origin.
One finds \cite{NatCommun.11.287} that an asymmetric oscillation
of the condensate can show up as a second frequency
in the gap oscillation below the well-known $2\Delta$ Higgs mode.
This asymmetric Higgs mode depends on the gap symmetry
and the respective asymmetric deviation.

We propose a phenomenological asymmetric driving scheme
to describe such effects in an experiment.
Due to a momentum dependent driving,
the gap symmetry gets altered by an additional symmetry component $f_k^A$.
We add such a term to the pseudomagnetic field
altering the gap symmetry with the same time-dependence
as the usual driving term
\begin{align}
    \vec b_k &= \VVV{-2\Delta'(f_\vk + \delta_A \sin^2(\Omega t)f_\vk^A)}
    {2\Delta''(f_\vk + \delta_A \sin^2(\Omega t)f_\vk^A)}
    {2\epsilon_\vk + \epsilon_\vk^A(t)}\label{eq:bk_asym}
\end{align}
where $\delta_A$ determines the strength of the asymmetric driving.
There is some similarity to \cite{PhysRevB.87.054503},
where, however static and ab-initio,
a composite pairing interaction leads to multiple Higgs modes
in the different pairing channels.
Here, we dynamically drive an asymmetric oscillation of the gap
to study the effect on the Higgs spectrum.

Our approach is purely phenomenological
and neglects additional polarization dependencies
which are likely to occur due to the asymmetric driving.
As the polarization dependence is hard to predict
without an in detail understanding of the actual microscopic coupling,
we neglect it for our approach.

Using the modified pseudomagnetic field
and performing a linearization in the same way as in Sec.~\ref{sec:gap_osci}
one would neglect important contributions
from the products between the asymmetric driving term
and the deviations, e.g. $\delta \sin^2(\Omega t)f_\vk^A x_\vk(t)$, etc.
Therefore, we solve the Bloch equations for this section numerically
without any approximation by integrating the differential equations in time.
As we haven't added an additional polarization dependence,
we can suppress the CDF contribution by choosing $\theta = \pi/4$
to obtain the contribution from the Higgs channel.

In Fig.~\ref{fig:thg_asym},
we show the temperature dependence of the THG intensity
for a $d$-wave gap with an asymmetric driving $f_\vk^A = 1$,
i.e. a distortion in the $s$-wave channel.
This corresponds to driving the $B_{1g}$ mode of the $d$-wave gap.
It does not corresponds directly to any experimental excitation scheme,
but acts as a proof of principle to study the induced effects.
Any possible experimental realizations are likely
a superposition of fundamental symmetries.
For low driving frequencies one can observe a second resonance peak
in the THG signal below the $2\Delta$ peak,
which is also accompanied by a phase change of $\pi/2$.
Higher driving frequencies do not show such a resonance.
This can be understood better, by calculating the frequency dependence
of THG for different temperatures
from which one can extract the temperature dependence of the modes.
For the chosen set of parameters,
the resulting curve of the second mode follows approximately
$\omega_{B_{1g}}(T) = 0.44\Delta(T)$
which is also shown in the upper row of Fig.~\ref{fig:thg_asym}.
As one can see, the chosen higher driving frequency of
$2\Omega = \Delta$ stays always above the second mode
which explains that there is no second resonance peak.
For the lower driving frequency of $2\Omega = 0.4\Delta$,
the second resonance peak appears at the point
where $2\Omega = \omega_{B_{1g}}(T)$

In a recent THG experiments on cuprates
a collective mode below $2\Delta$ was observed \cite{arxiv.1901.06675}.
An asymmetric driving and resonantly excitation of a $B_{1g}$ mode
(or other non-$A_{1g}$ modes) could be, in principle,
an explanation of additional modes.

\section{Summary and discussion}
\label{sec:conclusion}
In this work we analyze the induced gap oscillations
in unconventional superconductors
due to periodic driving with THz light
within BCS theory in the Anderson pseudospin formalism.
We show exemplary for $d$-wave
how a non $s$-wave symmetry broadens both the resonance peak
and the phase change in the oscillation amplitude
and derive analytic expressions for the oscillations.
The gap oscillations lead to third-harmonic generation
whose intensity shows a resonance peak
if twice the driving frequency coincidences
with the energy of the Higgs mode.
Although,
the calculated spectra for pure $s$- and $d$-wave differ significantly,
experimental conditions can wash out the spectrum for $s$-wave as well,
which makes a differentiation of the two gap symmetries
only from an observation of the spectrum difficult.
We compare contributions from CDF and Higgs
and illustrated that the Higgs contribution to THG
follows closely the amplitude of the gap oscillation,
whereas the CDF contribution shows still a sharp peak
even in the non $s$-wave case.
The temperature dependence of the THG signal
as measured in the experiment
shows the same characteristics as the THG signal
as a function of the driving frequency.

In addition one finds that more complex gap symmetries,
shown exemplary for $d$+$s$-wave,
which contain multiple local gap maxima show resonances for each maxima.
Hereby it is the contribution of the Higgs oscillation
which shows this feature as the composite gap contains two Higgs modes
with frequencies at these local gap maxima.
Thus,
THG experiments can be used to detect the occurrence of multiple Higgs modes
in unconventional superconductors.

For $d$-wave, or in general for gap symmetries
where the squared symmetry function is isotropic in $x$- and $y$- direction,
the polarization dependence of the THG signal is the same as for $s$-wave.
For a simple square-lattice tight-binding dispersion,
the Higgs contribution is polarization-independent,
whereas the CDF contribution has a characteristic dependency.
However, deviations both from the isotropy of the gap symmetry
as well as the simple tight-binding dispersion
lead also to deviations in the polarization dependence
and the Higgs contribution can become polarization dependent.
Yet, if the band-dispersion is known,
a measurement of the polarization dependency
allows to distinguish the individual contributions from CDF and Higgs.
This result is probably true even beyond a clean-limit analysis.
Typical impurity-scattering introduces no additional direction dependence
as the scattered momenta are randomly distributed.
Therefore, the polarization dependence should be unaffected
\cite{PhysRevB.99.224510,PhysRevB.96.020505}.

In our analysis for the clean-limit,
where the coupling to the vector potential
is exclusively through the non-parabolicity of the band dispersion,
the CDF contribution exceeds in general the Higgs contribution.
Nevertheless, this work gives an interesting insight
into the driven dynamics of unconventional superconductors.
On the one hand, the Higgs contribution can still show its unique features
as seen in the $d$+$s$-wave case with the two-peak structure,
which is a result from the Higgs contribution alone.
On the other hand as it was shown elsewhere
\cite{JPhysSocJpn.84.114711,JPhysSocJpn.87.024704,PhysRevB.96.155311,%
PhysRevB.99.224510,PhysRevB.99.224511},
effects beyond BCS theory and impurity scattering in the dirty-limit
can strongly enhance the Higgs contribution over the CDF contribution.

Finally, we propose an asymmetric driving scheme
to model experiments where the coupling of the driving field
acts non-symmetrically
with respect to the groundstate symmetry on the condensate.
Such asymmetry induces asymmetric oscillations of the condensate
which can show up as an additional oscillation frequency of the gap,
dependent on the symmetry of the gap and the deviation.
With our proof of principle calculation we show
that THG experiments should be in principle able to measure
asymmetric Higgs modes
and provide therefore the same information as pump-probe experiments.
However, as the proposed phenomenological asymmetric driving scheme
may be difficult to realize experimentally,
the information obtained by THG experiments are limited in this respect.

In addition to the measurement of the collective modes itself,
spectroscopy of Higgs modes may be used in future
as an alternative or complementary probe to investigate the gap symmetry.
A controlled excitation and observation of Higgs modes
allows to gain information about symmetry properties of the underlying gap
\cite{NatCommun.11.287}. In a more general sense,
THG experiments may also serve as a new measure
for defining superconductivity in nonequilibrium.
Recent experiments and theoretical studies on light-induced superconductivity
\cite{Science.331.189,PhysRevB.89.184516,Nature.7591.461,%
JPhysSocJap.88.044704,arxiv.1905.08638,arxiv.1908.10879}
raise the question on how one defines superconductivity
in a short-lived nonequilibrium state.
So far, the criteria only include the vanishing resistivity property
of superconductors measured by a divergent imaginary part
of the optical conductivity for $\omega\rightarrow 0$,
however the expelling of a magnetic field, i.e. the Meissner effect,
has not yet been considered.
As the Meissner effect is induced by the Anderson-Higgs mechanism,
a measurement of the Higgs mode
should be an equivalent fingerprint of superconductivity.
While the repulsion of a magnetic field on an ultrashort timescale
is difficult to measure or even impossible,
a resonant behavior of the THG signal
in the light-induced superconducting state
could potentially be realized.

In addition to pump-probe experiments,
where the gap is quenched by a short pulse
and the following intrinsic Higgs oscillations can be observed,
THG experiments can serve as an alternative tool
for identifying Higgs modes of a superconductor.
These driven experiments have some advantages
over the pump-probe experiments.
No ultra-short single-cycle pulses are required
and the strong-damping of Higgs modes in gaps with nodes
are partly overcome due to the forced periodic driving
and oscillation of the gap.
Thus, in the context of Higgs spectroscopy,
i.e. the detection and characterization of Higgs modes in superconductors,
THG experiments extend the range of possible experimental setups.

\begin{acknowledgments}
We thank B. Fauseweh, A. Schnyder and N. Tsuji
as well as S. Kaiser, H. Chu, M.J. Kim and R. Shimano
for helpful discussions both from theory and experimental side.
We also thank the Max Planck-UBC-UTokyo Center for Quantum Materials
for fruitful collaborations and financial support.
\end{acknowledgments}

\appendix

\section{Expansion of driving term}
\label{sec:expansion_driving_term}
We assume
that the $x$- and $y$-directions are equivalent
for $f_\vk^2$ and $\epsilon_\vk$,
i.e. $f(k_x,k_y)^2 = f(k_y,k_x)^2$
and $\epsilon(k_x,k_y) = \epsilon(k_y,k_x)$.
It follows for any function $a_\vk$
whose $\vk$-dependence stems only from $f_\vk^2$ and $\epsilon_\vk$ that
\begin{align}
    \sum_\vk (\partial_{xx}^2 \epsilon_\vk) a_\vk
    = \sum_\vk (\partial_{yy}^2 \epsilon_\vk) a_\vk
    = \frac 1 2 \sum_\vk (\nabla^2 \epsilon_\vk) a_\vk\,.
\end{align}
This assumption is true for functions like
$\epsilon_\vk = \epsilon( |\vec k|)$ or
$\epsilon_\vk \propto \cos k_x + \cos k_y$
and $f_\vk = 1$ or $f_\vk \propto \cos k_x - \cos k_y$.
As we are mostly interested in values close to the Fermi energy
$\epsilon_{\mathrm F}$, i.e. values of $\epsilon_\vk \approx 0$,
we expand the Laplacian of the dispersion to first order in the dispersion
\begin{align}
    \frac 1 2 \nabla^2 \epsilon_\vk
    = \alpha_0 + \alpha_1\epsilon_\vk + \mathcal O(\epsilon_\vk^2)
    \label{eq:laplace_expansion}\,.
\end{align}
As an example, for the single-band tight-binding dispersion
on the square lattice with nearest neighbor hopping~$t$
\begin{align}
    \epsilon_\vk = -2t(\cos k_x + \cos k_y) - \epsilon_{\mathrm F}
    \label{eq:sbnn_dispersion}
\end{align}
it follows
\begin{align}
    \frac 1 2 \nabla^2 \epsilon_\vk = t(\cos k_x + \cos k_y)
        = -\frac{\epsilon_{\mathrm F}}{2} - \frac{\epsilon_\vk}{2}
\end{align}
and the expansion to first order with $\alpha_0 = -\epsilon_{\mathrm F}/2$
and $\alpha_1 = - 1/2$ becomes exact.
Further we assume that the mixed derivatives of the dispersion vanishes,
i.e. $\partial_{xy}^2 \epsilon_\vk = 0$.

\section{Inverse Laplace transform}
\label{sec:inverse_laplace_transform}
\begin{figure}[t]
    \includegraphics[width=0.5\textwidth]{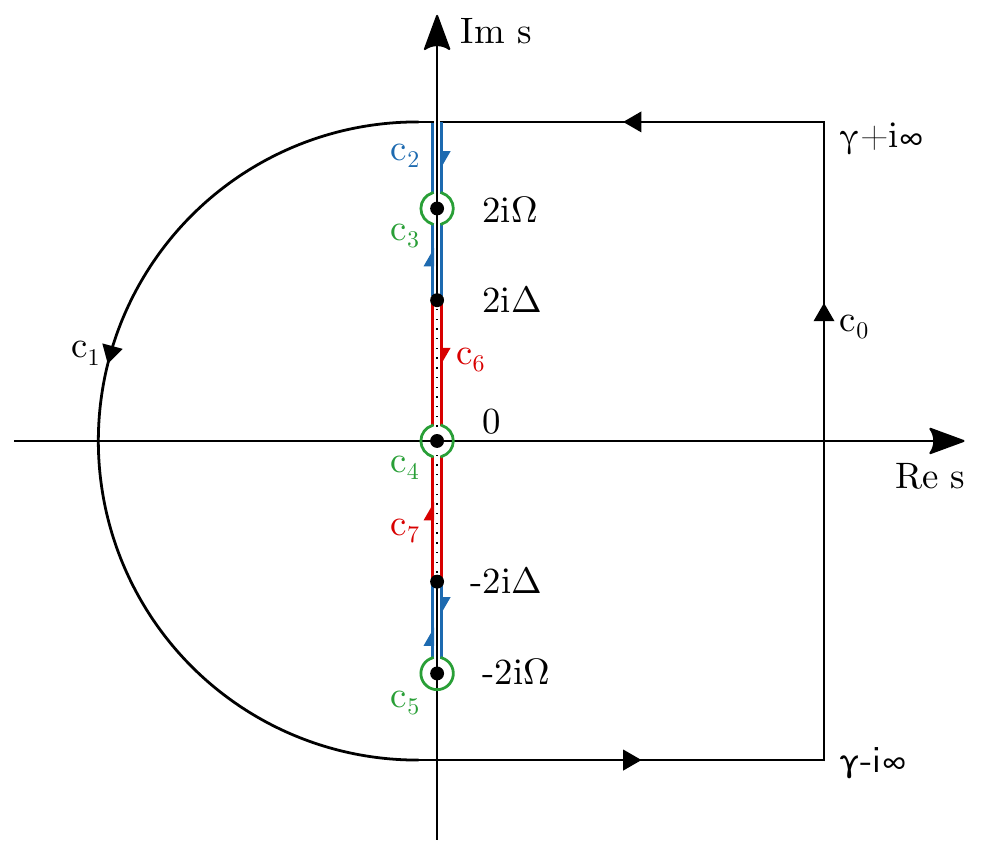}
    \caption{Poles at $s=0,\pm2\im\Omega$
    and continuous line of branch points at $s=2\im\Delta f$
    of the integrand in Eq.~\eqref{eq:I2}.
    The Bromwich integral $c_0$ is extended
    to the shown contour in the complex plane,
    where the residues of the poles $c_3$, $c_4$, $c_5$
    and the paths $c_6$, $c_7$ next to the branch cut contribute.}
    \label{fig:contour}
\end{figure}%
The inverse Laplace transform for $I_1(s)$ in Eq.~\eqref{eq:I1I2_I1}
is trivial and we get
\begin{align}
    I_1(t) &= \frac{1-\cos(2\Omega t)}{4}\,.
\end{align}
For the second term $I_2(t)$ in Eq.~\eqref{eq:I1I2_I2},
the Bromwich integral has to be explicitly evaluated
\begin{align}
    I_2(t) = \frac{1}{2\pi\im}\int_{\gamma-\im\infty}^{\gamma+\im\infty}
        \e^{st} I_2(s)\,\mathrm ds
    \label{eq:I2}\,.
\end{align}
The integrand has three poles at $s = 0,\pm 2\im\Omega$
and depending on $f \in [-1,1]$
a continuous line of branchpoints
between $s = -2\im\Delta$ and $s = 2\im\Delta$.
We evaluate the integral by extending the path into the complex plane
considering the contour in Fig.~\ref{fig:contour},
which is chosen such that the poles contribute with their residue
and a branch cut between $s = -2\im\Delta$ and $s = 2\im\Delta$
on the imaginary axis is excluded.
The closed loop integral vanishes as no poles are included,
the outer integral $c_1$ vanish if the radius goes to infinity.
The integral denoted by $c_2$,
i.e. the paths from infinity to the poles
and between the poles and the end of the branch cut
along the imaginary axis cancel each other.
We are left with the contributions around the poles $c_3$, $c_4$, $c_5$
and the paths left and right of of the branch cut $c_6$ and $c_7$
\begin{align}
    \int_{\gamma-\im\infty}^{\gamma+\im\infty}
    = - \int_{c_3} - \int_{c_4} - \int_{c_5} - \int_{c_6} - \int_{c_7}\,.
\end{align}
Each of the small circles around the poles
contribute with their residue in the limit of the radius going to zero
\begin{align}
    \sum_{n=3,4,5}\frac{1}{2\pi\im}\int_{c_n} \e^{st}I_2(s)\,\mathrm ds
    &= \sum_{r=0,\pm2\im\Omega}\operatorname{Res}_r(\e^{st}I_2(s))\,.
\end{align}
For the residues it follows
\begin{align}
    &\operatorname{Res}_0(\e^{st}I_2(s))
        = \frac{1}{4\lambda \int f^2 \,\mathrm d\varphi}\,,\\
    &\operatorname{Res}_{\pm2\im\Omega}(\e^{st}I_2(s))
        = -\frac{\Omega}{8}\e^{\pm2\im\Omega t}
        \notag\\&\qquad
        \times\frac{1}{\lambda\int\mathrm d\varphi\, f^2
            \sqrt{\Delta^2 f^2 -\Omega^2}
            \sin^{-1}\left(\frac{\Omega}{\Delta|f|}\right)
        }\,.
\end{align}
The integrals along $c_6$ and $c_7$ can be parametrized
with $s(r) = \epsilon + \im r$ with $r \in [2\Delta,-2\Delta]$ for $c_6$
and $r \in [-2\Delta,2\Delta]$ for $c_7$ for $\epsilon \rightarrow 0$.
We obtain
\begin{align}
    &\frac{1}{2\pi\im}\int_{c_4/c_6} \e^{st}I_2(s)\,\mathrm ds
    = \frac{\pm 1}{2\pi\im}\int_{-2\Delta}^{2\Delta}\mathrm dr\,
    \frac{\Omega^2}{4\Omega^2-r^2}
    \notag\\&\quad
    \times\frac{\e^{\pm\im rt}}{\lambda \int\mathrm d\varphi\, f^2
        \sqrt{4\Delta^2 f^2 - r^2}
        \sin^{-1}\left(\frac{r}{2\Delta|f|}\right)
    }\,.
\end{align}

\section{Derivation of THG}
\label{sec:deriv_thg}
The Fourier transform of Eq.~\eqref{eq:j_parallel} reads
\begin{align}
    j^{(3)}_{\parallel\perp}(\omega) &= - 2e^2
        \sum_\vk D_{\epsilon_\vk}^{\parallel\perp}(\theta)
        \notag\\&\qquad
        \times\frac{1}{\sqrt{2\pi}}
            \int A_0(\omega')z_\vk(\omega-\omega')\,\mathrm d\omega'\,.
\end{align}
The expression for the vector potential \eqref{eq:At} is inserted
and the convolution at $\omega = 3\Omega$ is evaluated
\begin{align}
    j^{(3)}_{\parallel\perp}(3\Omega) &= \im e^2 A_0
        \sum_\vk D_{\epsilon_\vk}^{\parallel\perp}(\theta)
            z_\vk(s=2\im\Omega)\,.
    \label{eq:j3_3w}
\end{align}
We make use of the solution $z_\vk(s=2\im\Omega)$
from the linearized Bloch equations \eqref{eq:zk_sol}
including also the temperature dependence.
Using the definitions
\begin{align}
    x_0(s) &= V \sum_\vk
        \frac{f_\vk^2}{E_\vk(4E_\vk^2 + s^2)}
        \tanh\left(\frac{E_\vk}{2k_BT}\right)\,,\label{eq:xsum0}\\
    x_1(s) &= V \sum_\vk
        \frac{f_\vk^4}{E_\vk(4E_\vk^2 + s^2)}
        \tanh\left(\frac{E_\vk}{2k_BT}\right)\,,\\
    x_2(s) &= V \sum_\vk
        \frac{\epsilon_\vk f_\vk^2}{E_\vk(4E_\vk^2 + s^2)}
        \tanh\left(\frac{E_\vk}{2k_BT}\right)\,,\\
    x_3(s) &= V \sum_\vk
        \frac{\epsilon_\vk^2 f_\vk^2}{E_\vk(4E_\vk^2 + s^2)}
        \tanh\left(\frac{E_\vk}{2k_BT}\right)\,,\\
    x_4^{\parallel\perp}(s) &= V \sum_\vk
        \frac{D_{\epsilon_\vk}^{\parallel\perp} f_\vk^2}{E_\vk(4E_\vk^2 + s^2)}
        \tanh\left(\frac{E_\vk}{2k_BT}\right)\,,\\
    x_5^{\parallel\perp}(s) &= V \sum_\vk
        \frac{\epsilon_\vk D_{\epsilon_\vk}^{\parallel\perp} f_\vk^2}
            {E_\vk(4E_\vk^2 + s^2)}
        \tanh\left(\frac{E_\vk}{2k_BT}\right)\,,\\
    x_6^{\parallel\perp}(s) &= V \sum_\vk
        \frac{D_{\epsilon_\vk}^\parallel D_{\epsilon_\vk}^{\parallel\perp}
        f_\vk^2}{E_\vk(4E_\vk^2 + s^2)}
        \tanh\left(\frac{E_\vk}{2k_BT}\right)\label{eq:xsum6}
\end{align}
one obtains
\begin{align}
    j_{\parallel\perp}^{(3)\mathrm{H}}(3\Omega) &\propto 2\Delta
        x_5^{\parallel\perp}(2\im\Omega) \delta\Delta'(2\im\Omega)
        \,,\\
    j_{\parallel\perp}^{(3)\mathrm{P}}(3\Omega) &\propto - s \Delta
        x_4^{\parallel\perp}(2\im\Omega)\delta\Delta''(2\im\Omega)
        \,,\\
    j_{\parallel\perp}^{(3)\mathrm{CDF}}(3\Omega)
        &\propto - \Delta^2 e^2A_0^2(2\im\Omega)
        x_6^{\parallel\perp}(2\im\Omega)
\end{align}
where the real and imaginary part of the gap can be written as
\begin{align}
    &\delta\Delta'(s) = \Delta e^2A_0^2(s)
        \notag\\&
        \times \frac{s^2 x_2(s)x_4^\parallel(s)
        + 2 x_5^\parallel(s)\Big(2\Delta^2 x_1(s) + 2x_3(s)-1\Big)
        }
        {2 s^2 x_2(s)^2
        + 2\Big(2 x_3(s)-1\Big)\Big(2\Delta^2x_1(s) + 2x_3(s) - 1\Big)
        }\\
    \intertext{and}
    &\delta\Delta''(s) = s \Delta e^2A_0^2(s)
        \notag\\&
        \times \frac{2x_3(s)x_4^\parallel(s) - x_4^\parallel(s)
            - 2 x_2(s) x_5^\parallel(s)
        }
        {2 s^2 x_2(s)^2
        + 2\Big(2 x_3(s)-1\Big)\Big(2\Delta^2x_1(s) + 2x_3(s) - 1\Big)
        }\,.
\end{align}
To understand these expression, we make use of the same approximations
as in the previous sections. It follows
\begin{align}
    x_2(s) &= 0\,,\\
    x_4^{\perp}(s) &= 0\,,\\
    x_5^{\perp}(s) &= 0\,,
\end{align}

\begin{align}
    x_4^{\parallel}(s) &= V \sum_\vk
        \frac{\partial_{xx}^2\epsilon_\vk f_\vk^2}{E_\vk(4E_\vk^2 + s^2)}
        \tanh\left(\frac{E_\vk}{2k_BT}\right)
        \notag\\
        &= \alpha_0 x_0(s)\,,\\
    x_5^{\parallel}(s) &= V \sum_\vk
        \frac{\epsilon_\vk \partial_{xx}^2\epsilon_\vk f_\vk^2}
            {E_\vk(4E_\vk^2 + s^2)}
        \tanh\left(\frac{E_\vk}{2k_BT}\right)
        \notag\\
        &= \alpha_1 x_3(s)\,,\\
    x_6^{\perp}(s) &= \frac 1 4 \sin 4\theta
        \notag\\&\quad
        \times V \sum_\vk
        \frac{(\partial_{xx}^2\epsilon_\vk)^2
        f_\vk^2}{E_\vk(4E_\vk^2 + s^2)}
        \tanh\left(\frac{E_\vk}{2k_BT}\right)
        \notag\\&\quad
        - \frac 1 4 \sin 4\theta
        \notag\\&\quad
        \times V \sum_\vk
        \frac{(\partial_{xx}^2\epsilon_\vk)(\partial_{yy}^2\epsilon_\vk)
        f_\vk^2}{E_\vk(4E_\vk^2 + s^2)}
        \tanh\left(\frac{E_\vk}{2k_BT}\right)
        \notag\\
        &\approx \frac 1 4\sin 4\theta
        \left(2\alpha_0^2 x_0(s) + 2\alpha_1^2 x_3(s)\right)\,,\\
    x_6^{\parallel}(s) &= \left(1-\frac 1 2 \sin^2 2\theta\right)
        \notag\\&\quad
        \times V \sum_\vk
        \frac{(\partial_{xx}^2\epsilon_\vk)^2
        f_\vk^2}{E_\vk(4E_\vk^2 + s^2)}
        \tanh\left(\frac{E_\vk}{2k_BT}\right)
        \notag\\&\quad
        +\frac 1 2 \sin^2 2\theta
        \notag\\&\quad
        \times V \sum_\vk
        \frac{(\partial_{xx}^2\epsilon_\vk)(\partial_{yy}^2\epsilon_\vk)
        f_\vk^2}{E_\vk(4E_\vk^2 + s^2)}
        \tanh\left(\frac{E_\vk}{2k_BT}\right)
        \notag\\
        &\approx \left(1-\frac 1 2 \sin^2 2\theta\right)
        \left(2\alpha_0^2 x_0(s) + 2\alpha_1^2 x_3(s)\right)\,.
\end{align}
where we identified
$\sin^4\theta + \cos^4\theta = 1-\frac 1 2\sin^2 2\theta$,
$2\sin^2\theta\cos^2\theta = \frac 1 2 \sin^2 2\theta$
and $\sin\theta\cos\theta(\cos^2\theta-\sin^2\theta) = \frac 1 4\sin 4\theta$
and neglected sums with terms
$\propto (\partial_{xx}^2\epsilon_\vk)(\partial_{yy}^2\epsilon_\vk)$.
We use the general fact that
$x_3(s) = \frac 1 2 - \Delta^2 x_1(s) - \frac 1 4 s^2 x_0(s)$
and the expression for the real part of the gap
reduces to the result in Eq.~\eqref{eq:dD1s} from the previous section
\begin{align}
    \delta\Delta'(s) &= \frac 1 2 \alpha_1 \Delta e^2 A_0^2(s)
        \left(1+\frac{1}{2x_3(s)-1}\right)
\end{align}
whereas the imaginary part reduces to Eq.~\eqref{eq:dD2s}
\begin{align}
    \delta\Delta''(s) &= -\alpha_0 \Delta e^2 \frac{A_0^2(s)}{s}\,.
\end{align}
Writing the sums as integrals
using $x_3(s) = 1/2 - H(s)/2$ and $x_0(s) = 2G(s)$
one obtains the result Eqs.~\eqref{eq:jHiggsApprox}-\eqref{eq:jCDFApprox}.

\end{document}